\documentclass[sigconf]{acmart}

\copyrightyear{2025}
\acmYear{2025}
\setcopyright{cc}
\setcctype{by}
\acmConference[CCS '25]{Proceedings of the 2025 ACM SIGSAC Conference on Computer and Communications Security}{October 13--17, 2025}{Taipei, Taiwan.}
\acmBooktitle{Proceedings of the 2025 ACM SIGSAC Conference on Computer and Communications Security (CCS '25), October 13--17, 2025, Taipei, Taiwan}
\acmDOI{10.1145/3719027.3765085}
\acmISBN{979-8-4007-1525-9/2025/10}

\usepackage{multirow} %
\usepackage{graphicx} %
\usepackage{hyperref} %
\usepackage{listings} %
\usepackage{array} %
\usepackage{amsmath} %
\usepackage{color} %
\usepackage[linesnumbered,ruled,vlined]{algorithm2e} %
\usepackage{soul} %
\usepackage{textcomp} %
\usepackage{bbding} %

\usepackage{amssymb} %

\usepackage{makecell}
\usepackage{enumitem}
\usepackage{tcolorbox}
\usepackage{pifont}
\usepackage{colortbl}
\usepackage{subcaption}
\usepackage{tikz}

\newcommand{\cmark}{\ding{51}}%
\newcommand{\xmark}{\ding{55}}%

\def\BibTeX{{\rm B\kern-.05em{\sc i\kern-.025em b}\kern-.08emT\kern-.1667em\lower.7ex\hbox{E}\kern-.125emX}}

\AtBeginDocument{%
  \providecommand\BibTeX{{%
    \normalfont B\kern-0.5em{\scshape i\kern-0.25em b}\kern-0.8em\TeX}}}

\newcommand{\cc}[1]{\mbox{\smaller[0.5]\texttt{#1}}}
\newcommand{\etal}{{\em et al.}\xspace}
\newcommand{\eg}{{\em e.g.,}\xspace}
\newcommand{\ie}{{\em i.e.,}\xspace}

\newcommand{\PP}[1]{\vspace{2px}\noindent{\bf#1.}\xspace}

\newcommand*\WC[1]{%
\begin{tikzpicture}[baseline=(C.base)]
\node[draw,circle,inner sep=0.2pt](C) {#1};
\end{tikzpicture}}

\newcommand{\boxbeg}{
\vspace{2px}
\noindent\begin{tabular}{|l|}\hline
\begin{minipage}{3.2in}
\vspace{2px}
\noindent
}

\newcommand{\boxend}{
\vspace{2px}
\end{minipage}\\ \hline
\end{tabular}
\vspace{-10pt}
}

\newcolumntype{P}[1]{>{\centering\arraybackslash}p{#1}}

\newtcolorbox{boxI}{
    colback = lightgray!10, 
    colframe = black, 
    boxrule = 0.5pt, 
    toprule = 0.5pt, %
    arc = 2pt,
    left = 1pt,
    right = 1pt,
    bottom = 0pt,
    top = 0pt
}

\newcounter{observcntr}
\newcommand*{\observ}[1]{%
    \stepcounter{observcntr}%
    \begin{center}
    \vspace{-2px}
        \begin{boxI}
        \textbf{Key Takeaway~\arabic{observcntr}: }{#1}.
        \end{boxI}
    \vspace{-8px}    
    \end{center}
}

\newcommand{\UP}[1]{\textcolor{black}{#1}}

\title{A Decade-long Landscape of Advanced Persistent Threats:\\
Longitudinal Analysis and Global Trends}

\author{Shakhzod Yuldoshkhujaev}
\affiliation{%
  \institution{Sungkyunkwan University}
  \city{Suwon}
  \country{Republic of Korea}
}
\email{shakzod02@g.skku.edu}

\author{Mijin Jeon}
\affiliation{
  \institution{Sungkyunkwan University}
  \city{Suwon}
  \country{Republic of Korea}
}
\email{jinijxxn@g.skku.edu}

\author{Doowon Kim}
\affiliation{
  \institution{University of Tennessee}
  \city{Knoxville}
  \state{TN}
  \country{United States}
}
\email{doowon@utk.edu}

\author{Nick Nikiforakis}
\affiliation{
  \institution{Stony Brook University}
  \city{Stony Brook}
  \state{NY}
  \country{United States}
}
\email{nick@cs.stonybrook.edu}

\author{Hyungjoon Koo}
\authornote{Corresponding author.}
\affiliation{
  \institution{Sungkyunkwan University}
  \city{Suwon}
  \country{Republic of Korea}
}
\email{kevin.koo@skku.edu}

\renewcommand{\shortauthors}{Shakhzod Yuldoshkhujaev, Mijin Jeon, Doowon Kim, Nick Nikiforakis, and Hyungjoon Koo}
\begin{document}

\begin{abstract}
An advanced persistent threat (APT) refers to
a covert and long-term cyberattack, 
typically conducted by state-sponsored actors, 
targeting critical sectors and often remaining 
undetected for long periods. 
In response, collective intelligence 
from around the globe collaborates 
to identify and trace surreptitious activities, 
generating substantial documentation 
on APT campaigns publicly available on the web.
While
a multitude of prior works predominantly 
focus on specific aspects of APT
cases, such as 
detection, evaluation, 
cyber threat intelligence, and dataset creation,
limited attention
has been devoted to revisiting and 
investigating these scattered
dossiers in a longitudinal manner. %

The objective of our study lies 
in filling the gap by offering
a macro perspective,
connecting key insights and global trends 
in the past APT attacks.
We systematically analyze
six reliable sources---
three focused on technical reports
and another three on threat actors---
examining 1,509 APT dossiers 
(\ie totaling 24,215 pages) spanning
from 2014 to 2023 (a decade), and 
identifying 603 unique APT groups
in the world.
To efficiently unearth relevant information, 
we employ a hybrid methodology that combines
rule-based information retrieval with
large-language-model-based search techniques.
Our longitudinal analysis reveals shifts 
in  threat actor activities, 
global attack vectors,
changes in targeted sectors, and 
the relationships between cyberattacks 
and significant events, 
such as elections or wars,
which provides insights 
into historical patterns in APT evolution. 
Over the past decade, 154 countries 
have been affected, primarily 
using malicious documents
and spear phishing as 
the dominant initial infiltration vectors,
and a noticeable decline in zero-day 
exploitation since 2016. 
Furthermore, we present our findings
through interactive visualization
tools, such as an APT map 
or a flow diagram, to facilitate 
intuitive understanding of
the global patterns and trends in
APT activities.
\end{abstract}

\ccsdesc[500]{General and reference~Measurement}
\ccsdesc[300]{Security and privacy~Social aspects of security and privacy}

\keywords{Advanced Persistent Threats; Longitudinal Analysis; Global Trends}

\maketitle

\section{Introduction}
\label{sec:intro}
Advanced persistent threats (APTs) are covert 
and sophisticated cyberattacks, 
typically orchestrated by state actors.
APT campaigns attempt to gain unauthorized access to 
remote machines and stay undetected 
for extended periods, enabling targeted 
campaigns against %
governments and financial 
institutions~\cite{APT_new_survey}. 
Their primary objectives are 
to steal sensitive data, 
to disrupt critical operations, and 
to undermine national security or 
economic stability~\cite{APT_survey}.

Given the severity of risks and threats posed by APTs, both industry and academia put in constant 
effort to monitor and understand APT-involving incidents. 
The security industry 
(\eg companies, experts, practitioners) 
investigates each APT incident, 
producing invaluable 
technical reports and articles. 
These \emph{scattered} dossiers from varying
sources provide detailed information 
on attack strategies, threat actors, 
and exploitation techniques on the web.
Moreover, security practitioners track  
technical reports~\cite{TR1, TR2, TR3} 
and threat actors~\cite{MISP, EternalLiberty, APTmap_3d_git}, 
organizing them
in (public) repositories.
The security industry also compiles Cyber Threat Intelligence (CTI)~\cite{cti}
databases to detect, analyze,
and mitigate threats, including
\WC{1} Indicators of Compromise 
(IoCs)~\cite{IoCs} for 
collecting forensic evidence, 
\WC{2} Common Vulnerabilities
and Exposures (CVEs)~\cite{CVEs} for maintaining 
publicly disclosed security flaws, 
\WC{3} MITRE ATT\&CK Techniques~\cite{MITRE_ATTaCK} 
for identifying adversarial tactics, 
and \WC{4} YARA rules~\cite{Yara}
for detecting specific malware patterns.
For the intuitive understanding of APTs,
visual representations with a 
map~\cite{kaspersky_cybermap, 
aptmap_netlify, APTmap_3d_git}
have been implemented on the web.
Orthogonally to the aforementioned work by the cyber-security industry, academics also invest resources in better understanding past APTs.
The majority of recent academic studies have focused on studying and analyzing specific attributes related to APT incidents, such as detection and 
evaluation~\cite{ModelnDetection, ProvG, GradientRules, highVolumeNetwork, holmes_apt, magic_apt, Wang_2024, Malik2024Advanced, TacticalProvenance, mitre_edr_performance}, 
CTI~\cite{aCTIon, apt_behind_scene, ttpHunter, cyber-kill, semantic_ranking, trec}, and APT dataset~\cite{aCTIon, trec, cyber-kill, semantic_ranking, APT_datasets}.

In this paper, we observe that
few studies have systematically 
investigated the landscape of APTs 
over an extended period of time.
To help draw a complete picture of APT activity,
we study the landscape of APT incidents over a period of 10 years.
We uncover longitudinal changes
through an in-depth analysis of fragmented, publicly available
APT documentation.
Our analysis can assist in identifying
broader trends and patterns in surreptitious 
APT activities, offering valuable insights 
into the evolution of APT targets, malware samples, and sophisticated attack techniques.

More specifically, this study attempts to identify global trends 
from a macro viewpoint, including
vulnerability exploitations, 
threat actor behaviors, and 
target changes through 
a large-scale investigation of
$1,509$ unique APT dossiers and 
$603$ APT groups 
\emph{over the past decade}.
Due to the large volume of publicly available APT reports
(\ie $24,215$ pages of technical reports), 
we adopt a hybrid 
information-retrieval approach by 
leveraging the inference capabilities
of Large Language Models (LLMs)
combined with a rule-based extraction tool.
To boost the LLM's accuracy, 
we carefully design 
the questions and prompts, evaluating 
multiple models to identify 
the most effective retrieval for our goals.

With a comprehensive collection of APT dossiers~\cite{TR1, TR2, TR3},
our study aims to analyze
\WC{1} the evolution of APTs over the past decade
in terms of victim countries, threat actors, 
target sectors, initial attack vectors, and zero-day vulnerabilities;
\WC{2} CTI records in APT cases 
(\eg CVEs, MITRE ATT\&CK, YARA rules);
\WC{3} common traits of APT campaigns
that demonstrate concealment (\eg attack duration) and aggressiveness; and
\WC{4} external factors affecting APT campaigns,
such as political events, international conflicts, 
global pandemics, or economic instabilities.

Our analysis yielded the following findings: 
\WC{1} Over the past decade, 
APT campaigns have impacted 154 countries
(80\% of all nations), with the United States, 
India, and South Korea among the most 
frequently targeted. 
While 
$446$ unique threat actors %
have been identified in our dataset, 
a small set of actors is responsible 
for a significant share of attacks, including
Lazarus~\cite{LazarusDescr}, APT28~\cite{APT28Descr}, 
and APT29~\cite{APT29Descr}. 
In terms of targets, the government and corporate sector 
attract the majority of APT attention, with malicious documents 
and spear phishing serving as the 
dominant initial infiltration vectors. 
\WC{2} Vulnerability-wise, while the exploited CVEs are highly severe (average score of $8.5$),
our findings
indicate that many of the attacks do not need to rely on zero-day vulnerabilities to be successful, which
peaked between 2014 and 2016 but has declined
thereafter. 
\WC{3} In terms of the lifetime of the recorded incidents, APT campaign duration varies widely, 
from a single day to nearly five years 
(137 days on average).
\WC{4} Finally, an important finding is that APT activity frequently coincided 
with political events, international conflicts, 
global crises like COVID-19, indicating that attackers had already performed
target reconnaissance and were waiting for an opportune time to act upon their findings.

To facilitate the exploration of APT 
campaign data by reviewers and eventually the general public, we designed an interactive 
map~\footnote{\url{https://lngt-apt-study-map.vercel.app/}} 
that incorporates decade-long historical data, 
including threat actor(s), CVEs, 
attack vector(s), malware, target sector(s), 
and estimated duration, with support 
for \emph{dynamic updates} using LLMs 
to retrieve content from technical reports. 
Additionally, we provide a flow
diagram~\footnote{\url{https://public.tableau.com/views/TopMentionedCountries/Top30Countries}} 
illustrating the relationships between 
selected threat actors and target countries. 

\noindent In summary, this paper makes the following original contributions:
\begin{itemize}[leftmargin=*]
    \item We conduct a longitudinal 
    measurement study of APT campaigns 
    over the last decade (2014 -- 2023), 
    organizing $1,509$ unique 
    technical reports, $603$ threat actors, 
    and $177$ news articles.
    \item To unveil longitudinal APT campaigns, 
    we retrieve and refine
    responses to 10 identified questions
    using (context-aware) LLMs.
    \item We carefully define four research 
    questions centered on the evolution,
    CTI records, common traits, and 
    external dynamics of APT campaigns
    over a decade.
    Our findings reveal global trends
    and key insights, including that
    the campaigns have affected 
    80\% of countries worldwide; 
    a small number of actors are responsible 
    for a disproportionate share of attacks; and
    the exploitation of both zero-day and one-day vulnerabilities is prevalent.
    \item We publicly release our curated dataset~\footnote{\url{https://zenodo.org/records/16869733}} 
    and an interactive APT-campaign map to foster
    future research in the field of APT studies.

\end{itemize}

\section{Background}
\label{sec:bg}

This section provides some background on
advanced persistent threats, CTI, and the focus of prior work on APT
campaigns.

\begin{table}[t!]
    \centering
    \caption{Summary of prior work on APTs. 
    We classify them into five categories
    where the majority of those studies
    focus on specific domains, covering
    limited periods.
    Our work aims to offer insights
    into the evolution of APTs over 
    the past decade
    from macro perspective (Section~\ref{sec:bg}).
    }
    \resizebox{0.99\linewidth}{!}{
    \begin{tabular}{>{\raggedright\arraybackslash}m{4cm}m{6.5cm}}
        \toprule

\textbf{Category} & \textbf{Topic Focus} \\ \midrule
\textbf{Survey} & 
APT Survey~\cite{APT_survey, APT_new_survey} \\ \midrule
\raggedright\textbf{Detection and evaluation} & \makecell[l]{
Detection techniques~\cite{ModelnDetection, ProvG, GradientRules, highVolumeNetwork, holmes_apt, magic_apt, Wang_2024, Malik2024Advanced}  \\
Evaluating APT detection systems~\cite{Malik2024Advanced} \\ 
APT reconstruction~\cite{TacticalProvenance, mitre_edr_performance}} \\ \midrule
\textbf{Cyber Threat Intelligence} & \makecell[l]{
Information retrieval~\cite{aCTIon, apt_behind_scene, ttpHunter, cyber-kill} \\
Information recognition~\cite{semantic_ranking, trec}} \\ \midrule 
\textbf{APT Dataset} & \makecell[l]{Dataset creation~\cite{aCTIon, trec, cyber-kill, semantic_ranking}\\
Dataset evaluation~\cite{APT_datasets}} \\ \midrule
\textbf{Technical articles} & 
\makecell[l]{APT trends~\cite{kasper_apt_trends,cybelangel} \\
Special reports~\cite{cisa_iran_apt, apt29_german_politics} }\\
\bottomrule

    \end{tabular}
    }
    \label{tab:priorwork}
\end{table}

\PP{Advanced Persistent Threats} 
APTs refer to a critical and insidious 
category of cyberattacks characterized by 
their sustained, targeted, and highly 
sophisticated nature. 
Typical attacks include individual actors attacking any and all systems for financial gain, activism, or to demonstrate technological proficiency.
However, APTs are often
executed by well-organized, resource-rich 
groups with strategic 
objectives~{\cite{APT_survey}}. 
These entities persist in their efforts to infiltrate and maintain a presence within a network over extended periods, often evading detection.
Evidenced by exploiting zero-day 
vulnerabilities, deploying custom-developed 
malware, and adopting advanced social-engineering techniques, APTs'
resource-intensive operations indicate
clear underlying motives - 
political~\cite{apt29_german_politics, political_china_cyber_attack_2024, political_macron_leaks_2017}, economic~\cite{economic_carderbee_certificate_abuse_2024, economic_inpage_zero_day_2016, economic_kawasaki_cyber_attack_2020}, and military~\cite{military_china_cyber_offensive_2024, military_positive_technologies_2024, military_us_treasury_2024}. 
APT campaigns predominately aim at espionage,
disruption, or sabotage, posing serious
threats to national security and critical
infrastructure. 

\PP{Cyber Threat Intelligence}
Cyber Threat Intelligence~\cite{cti} is 
an essential pillar of 
modern cybersecurity for threat
detection, prevention, and response. 
CTI enables organizations to collect 
and analyze information on cyber threats, 
anticipate emerging threats,
and mitigate potential risks beforehand, 
empowering them to take proactive
measures against cyberattacks.
Indicators of Compromises~\cite{IoCs} are 
critical elements of CTI, 
representing specific pieces of evidence 
that signal malicious activities within a 
system or network. 
Integrating such artifacts with CTI 
serves as actionable data for 
detecting and preventing future threats. 
IoCs are often aligned with standardized 
CTI frameworks to maximize 
their effectiveness.
For instance, the MITRE 
ATT\&CK~\cite{MITRE_ATTaCK} framework
categorizes adversarial behaviors 
into distinct techniques, 
offering a structured approach to 
understanding how IoCs align 
within the attack lifecycle. 
In a similar vein, CVEs~\cite{CVEs}
maintain up-to-date
(publicly known) vulnerabilities 
exploited by attackers, while 
YARA rules~\cite{Yara} 
facilitate the detection of malicious artifacts through signature-based matching.
As such, various tools have been developed 
to automate the extraction of IoCs 
from technical reports. 
One notable example is 
IoCParser~\cite{iocparser}, which 
specializes in retrieving diverse IoCs from URLs and texts. 
We leverage IoCParser to extract 
three primary IoCs -- 
CVEs, MITRE ATT\&CK technique IDs, and YARA rules.

\PP{Previous APT Studies}
\autoref{tab:priorwork} summarizes prior research on APTs. 
While extensive surveys on APTs have been 
conducted~\cite{APT_survey, APT_new_survey}, 
the majority of these studies are dedicated to specific domains
or cover limited periods. 
Beyond those surveys, we classify existing work into four distinct areas.
First, a significant portion of 
the literature concentrates on
detection and evaluation by exploring 
detection techniques~\cite{ModelnDetection, 
ProvG, GradientRules, highVolumeNetwork, 
holmes_apt, magic_apt, Wang_2024, 
Malik2024Advanced}, evaluating existing 
detection systems~\cite{Malik2024Advanced}, or 
reconstructing APT 
scenarios~\cite{TacticalProvenance, 
mitre_edr_performance}.
Second, several studies focus on information 
retrieval~\cite{aCTIon, apt_behind_scene, 
ttpHunter, cyber-kill} and information 
recognition~\cite {semantic_ranking, trec}
contributing to a deeper understanding of 
Cyber Threat Intelligence.
Third, another direction explores 
APT datasets in terms of their 
creation~\cite{trec, cyber-kill, 
semantic_ranking} and evaluation~\cite{APT_datasets}.
Lastly, the security industry has analyzed
APT trends~\cite{kasper_apt_trends, cybelangel} 
and issued special reports on particular 
incidents~\cite{cisa_iran_apt, apt29_german_politics}.
However, our work differs from prior work 
by offering a longitudinal analysis
based on a multitude of scattered APT dossiers, 
providing valuable insights into 
the evolution of APTs over time.

\begin{figure}[t!]
    \centering
    \resizebox{0.99\linewidth}{!}{
    \includegraphics{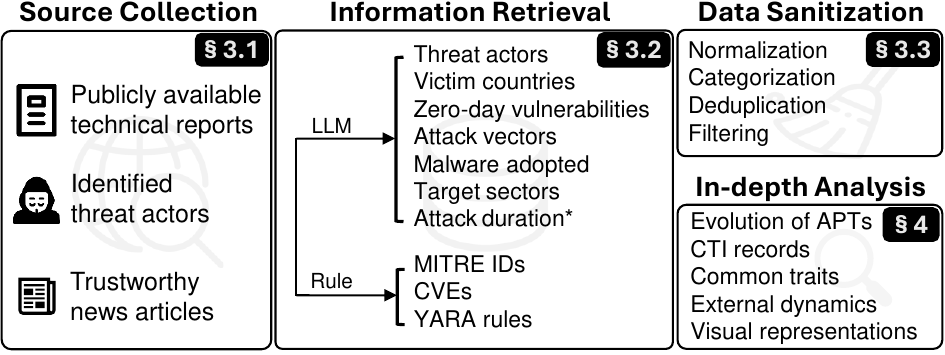}
    }
    \caption{Overview of 
    our methodology for 
    longitudinal APT analysis.
    We collect technical
    reports, threat actors,
    and news articles across
    the web (Section~\ref{sec:dataset}).
     Then, we probe valuable
     information from technical reports 
     based on rules and
     LLMs (Section~\ref{ss:retrieval}).
     Note that we manually inspect
     attack duration (*)
     for precise analysis.
     Next, we refine raw 
     information via normalization, 
     categorization, de-duplication,
     and filtering 
     (Section~\ref{subsec:sanitize}).
     Lastly, we conduct in-depth
     analyses to answer 
     our research questions
     (Section~\ref{sec:eval}). 
     }
    \label{fig:methodology}
\end{figure}

\PP{Threat Maps} 
Geo-location maps are often used to visualize APT incidents.
Kaspersky~\cite{kaspersky_cybermap}, for example, offers a cyber attack map that depicts real-time worldwide cyber assaults. 
However, this tool needs to include historical data for APT activities. 
Both APT threat actor map~\cite{aptmap_netlify} and APTMAP~\cite{APTmap_3d_git} aim to 
map APT-related data. 
However, both focus on APT groups' physical locations, which are missing a longitudinal study. 
Although these works introduce 
an APT map, they focus on
the distribution of victim nations by 
year alone. 
In this work, we devise a new APT map 
that represents combined information 
associated with
a certain threat actor or victim country
(\autoref{fig:aptMap} in Appendix).

\section{Methodology}
\label{sec:methodology}

This section sketches our methodology 
to reveal the landscape of APT campaigns
over the past 10 years, including details on
source collection, information retrieval,
and raw data refinement.

\PP{Overview}
\autoref{fig:methodology} 
illustrates
the overview of our methodology to
analyze decade-long APT cases 
spanning from $2014$ to $2023$.
We explore open APT technical reports that 
provide details of an individual attack, 
threat actors, and varying
news articles related to APT attacks (Section~\ref{sec:dataset}).
Among those sources, we choose the dossiers 
that are \WC{1} publicly accessible, 
\WC{2} offering up-to-date information, 
and \WC{3} written by a trustworthy 
entity.
Additionally, we gather past articles
from an authoritative outlet 
specializing in security-focused news.
Considering the volume of technical reports 
(\ie $24,215$ pages),
we retrieve useful information
(\eg victim countries, target sectors,
attack vectors)
by leveraging an LLM that helps contextual
inference.
Note that we also utilize a 
rule-based (\ie regular expressions) tool
that performs better than the LLM probe when looking for specific
information, such as, MITRE IDs, CVEs, and YARA rules.
(Section~\ref{ss:retrieval}).
Subsequently, we refine the 
raw data using normalization,
categorization, deduplication, and filtering
(Section~\ref{subsec:sanitize}).

\subsection{APT Source Collection}
\label{sec:dataset}

\begin{table}[t!]
    \centering
    \caption{Statistics on
    our collection of technical
    reports (TRs) and news articles on APTs.
    Out of 2,563 TRs,
    we analyze $1,509$ unique TRs (after removing $1,003$ duplicates and $51$ APT trend dossiers), along with $177$ news articles. 
    The numbers in parentheses indicate 
    the sum of all TRs before refinement.
    We discuss the credibility of 
    TR sources in Section~\ref{sec:dataset}.
    }
    \resizebox{0.99\linewidth}{!}{
    \begin{tabular}{lrrrrr}
        \toprule

\textbf{Year} & \textbf{TR\#1}~\cite{TR1} & \textbf{TR\#2}~\cite{TR2} & \textbf{TR\#3}~\cite{TR3} & \textbf{All TRs} & \textbf{News Articles} \\ [0.5ex] \midrule
        \textbf{2014} & 128 & 104 & 16 & 124 (248)& 2\\
        \textbf{2015} & 150 & 87 & 28 & 135 (265)& 8\\
        \textbf{2016} & 171 & 104 & 13 & 168 (288)& 14\\
        \textbf{2017} & 124 & 87 & 26 & 142 (237)& 15\\
        \textbf{2018} & 169 & 24 & 46 & 160 (239)& 18\\
        \textbf{2019} & 222 & 27 & 57 & 201 (306)& 15\\
        \textbf{2020} & 226 & 10 & 105 & 207 (341)& 21\\
        \textbf{2021} & 156 & 13 & 94 & 162 (263)& 19\\
        \textbf{2022} & 51 & 77 & 117 & 136 (245)& 30\\
        \textbf{2023} & 21 & 42 & 68 & 74 (131)& 35\\ 
        \midrule
\textbf{Total} & 1,418 & 575 & 570 & \textbf{1,509 (2,563)} & \textbf{177}\\
\bottomrule

    \end{tabular}
    }
    \label{tab:articles_stats}
\end{table}

As a primary source, we use 
a series of open datasets that 
collect technical reports and threat actors.
We acknowledge the collective intelligence of 
security experts and practitioners
who collaborate to uncover covert 
threat activities around the globe.
Of all, we carefully select reliable sources, 
which focus on three collections of
technical reports~\cite{TR1,TR2,TR3}
(\autoref{tab:articles_stats}) and 
another three collections of threat 
actors~\cite{MISP, EternalLiberty, 
APTmap_3d_git} (\autoref{tab:threat_stats}).
While the former provide in-depth 
analyses of each APT incident,
the latter offer individual APT group 
information.
We include both sources because 
technical reports often lack
detailed information about APT groups.

\begin{table}[t!]
    \centering
    \caption{Statistics on our
    collection of threat actors (TAs).
    Out of 1,684 TAs, 
    we organize 603 unique TAs 
    after removing 800 duplicates 
    and 281 entries containing 
    no information beyond their names.
    Note that the information of APT
    groups has been maintained by
    reliable sources
    (Section~\ref{sec:dataset}).
    }
    \resizebox{0.99\linewidth}{!}{
        \begin{tabular}{llr}
             \toprule
             
\textbf{Collection} &\textbf{Source of Threat Actors' Information} & \textbf{\# of TAs} \\ [0.5ex] \midrule
        \makecell[t]{\textbf{TA\#1}~\cite{MISP}} & \makecell[l]{MISP Project~\cite{wagner2016misp}} & 562 \\
        \midrule
        \makecell[t]{\textbf{TA\#2}~\cite{EternalLiberty}} & \makecell[l]{Palo Alto~\cite{paloalto}, 
        IBM X-Force\cite{IBM_hive, IBM_itg}, \\
        Malpedia~\cite{CybIN}, 
        Kaspersky~\cite{Kasper_APT}, 
        Crowdstrike~\cite{CrowdStrike},\\ 
        Mandiant~\cite{mandiant}, 
        Secureworks~\cite{secureWorks}, 
        Dragos~\cite{dragos}, \\
        Venafi~\cite{venafi}, 
        CERT-UA~\cite{cert-ua}, 
        Microsoft~\cite{microsoft}} & 692 \\ \midrule
        \makecell[t]{\textbf{TA\#3}~\cite{APTmap_3d_git}} & \makecell[l]{
        MITRE ATT\&CK~\cite{MITRE_ATTaCK}, 
        ETDA~\cite{ETDA}, \\
        VX-underground~\cite{VX_under}} & 430 \\ \midrule
        \textbf{Total} & -- & \textbf{603 (1,684)} \\ 
\bottomrule

        \end{tabular}
    }
    \label{tab:threat_stats}
\end{table}

\PP{Collection of Technical Reports}
We thoroughly examined the available web 
sources to obtain the most reputable 
technical reports (TRs), which 
we refer to as dossiers.
We adopt three TRs as reliable sources in this work. 
First, TR\#1~\cite{TR1} is a GitHub 
repository containing an extensive 
collection of $1,418$ TRs.
This repository is continually updated 
as new dossiers on APT cases become available.
Similarly, TR\#2~\cite{TR2} is
another GitHub repository that hosts 
$575$ TRs documenting 
various APT campaigns.
Lastly, TR\#3~\cite{TR3} is a curated list
of $570$ TRs included as
part of the comprehensive dataset 
maintained by Malpedia~\cite{CybIN}.
As shown in \autoref{tab:articles_stats}, 
we aggregate these three sources
as \emph{the collection of $1,509$ technical reports} between 2014 and 2023 
for our longitudinal analysis of APT cases.
Although a small number of 
old TRs
are available, we define
our collection from the year
of 2014 that shows a significant rise
in the volume of TRs.
Similarly, we exclude 2024 due to the small number of available TRs.
 Finally, we convert all
 web page reports into 
 the PDF format for consistency.

 \begin{figure}[t!]
    \centering
    \resizebox{0.85\linewidth}{!}{
        \includegraphics[trim=10 10 10 10, clip, width=1\textwidth]{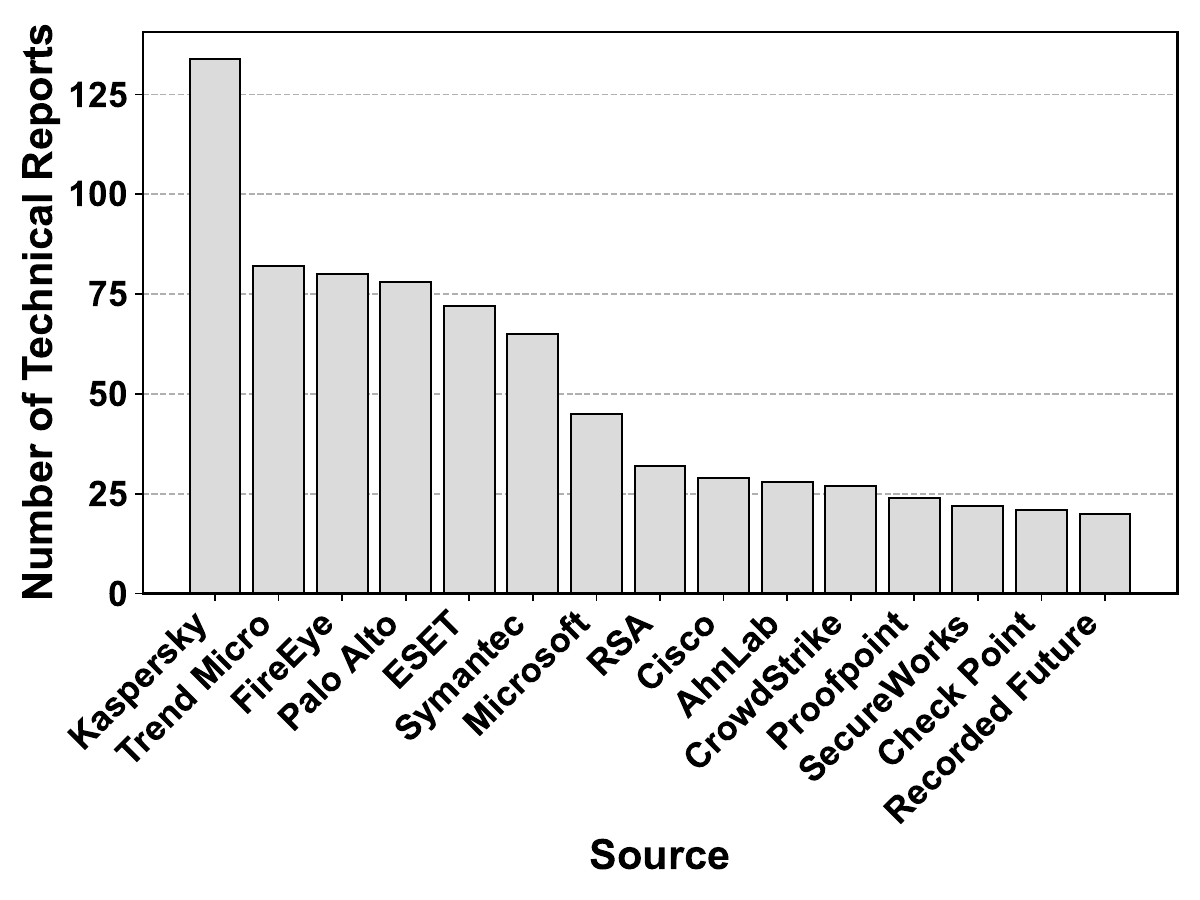}
    }
    \caption{Top 15 sources 
    from the collection of technical reports.
    Most reports come from
    reputable sources such as 
    Kaspersky~\cite{kasperskyInfo} and 
    Trend Micro~\cite{TrendMicroDescr}. 
    We confirmed that 
    1,412 (93.6\%) TRs are highly credible
    (Section~\ref{sec:dataset}).
    }
    \label{fig:top_sources}
\end{figure}

\PP{Collection of Threat Actors (APT Groups)}
We separately collect threat actors (TAs) since the collection of TRs does not contain that information.
First, TA\#1~\cite{MISP} 
includes information about the threat 
actors obtained from the MISP project~\cite{wagner2016misp},
one of the well-known Open Source Threat Intelligence Platforms.
Second, TA\#2~\cite{EternalLiberty} is
maintained by diverse security
companies and non-profit organizations
which includes a wide range of APT groups.
Third, TA\#3~\cite{APTmap_3d_git} updates
APT group information independently from 
the above two sources.
Similar to the collection
of TRs, we combine all the above
three sources, 
obtaining \emph{the collection
of $603$ unique threat actors}.
\autoref{tab:threat_stats} provides the 
statistics of our collection, which
encompasses the information on 
TAs' official names and their alias(es), country of origin, motives, first-seen year, activity patterns, and sponsors.

\PP{Collection of News Articles}
Oftentimes, APTs occur in the context of cyber warfare.
To identify plausible connections between APTs 
and external factors (including 
national conflicts, geopolitical events, 
and global crises)
we gathered security news articles 
and media reports
about APT incidents for the last decade. 
Note that our collection includes $177$ 
news articles (\autoref{tab:articles_stats}) 
through the Google News search 
engine~\cite{googleNews} using 
the \cc{APT} and \cc{attack} keywords.

\PP{Source Credibility}
\label{reliability}
Characterizing the sources is crucial 
for further analysis since
unreliable source(s) could severely 
distort our collected data,
compromising the validity of the findings. 
First, we explicitly extract 
the origin of a report 
(\eg organization name)
from the collection of TRs.
We confirmed that $1,412$ ($93.6\%$)  of 
our TR collection come 
from highly reliable sources, 
including reputable security companies 
like Kaspersky~\cite{kasperskyInfo}, 
RSA~\cite{rsaInfo}, 
FireEye~\cite{fireeyeInfo}, and 
Microsoft~\cite{microsoftInfo}, 
as well as governmental agencies 
like NATO~\cite{natoInfo}, 
US-CERT~\cite{usCertInfo}, FBI~\cite{fbiInfo}, 
or trustworthy news sources. 
\autoref{fig:top_sources} highlights the 
top $15$ sources, collectively representing 
approximately half of TRs. 
The remaining $97$ TRs ($6.4\%$)
come from individual web blogs 
and security experts, which may be 
relatively less reputable or credible.
Second, for the collection of TAs,
\autoref{tab:threat_stats} shows that 
the original data comes from highly trusted 
companies and projects that deal with CTI, 
such as MITRE~ATT\&CK~\cite{MITRE_ATTaCK}, 
MISP~\cite{wagner2016misp}, ETDA~\cite{ETDA}, and 
CrowdStrike~\cite{CrowdStrike}.

\subsection{APT Information Retrieval}
\label{ss:retrieval}

We carefully define ten items 
that have the potential to reveal longitudinal APT
changes and global trends, including 
MITRE IDs~\cite{MITRE_ATTaCK}, 
CVEs~\cite{CVEs},
YARA {rules}~\cite{Yara},  
threat actors (APT groups), 
victim country, zero-day vulnerabilities,
initial attack vectors, associated malware, 
target sectors, and attack durations.

\subsubsection{LLM-based Retrieval}
\label{subsec:LLMMethod}

In this work, we leverage 
the language model’s contextual 
inference capabilities to accurately extract 
relevant information from 
our collected technical reports.

\begin{table}[t!]
    \centering
    \caption{
    Performance comparison of precision (P), 
    recall (R), and F1
    scores across three different LLMs. 
    To evaluate LLM models, we obtain 
    the ground truth by manually 
    inspecting around 120 technical reports.
    Due to its superior performance, 
    we choose the GPT-4-Turbo model
    (Section~\ref{subsec:LLMMethod}).
    }
    \resizebox{1\linewidth}{!}{
    \begin{tabular}{lrrrrrrrrr}
        \toprule

\textbf{Language Model} & \multicolumn{3}{c}{\textbf{Gemini Flash~\cite{geminiFlash}}} & \multicolumn{3}{c}{\textbf{GPT-4-Turbo~\cite{gpt4}}} & \multicolumn{3}{c}{\textbf{GPT-4o~\cite{gpt4o}}} \\
\textbf{Search Item} & \textbf{P} & \textbf{R} & \textbf{F1} & \textbf{P} & \textbf{R} & \textbf{F1} & \textbf{P} & \textbf{R} & \textbf{F1} \\ [0.5ex]\midrule
\textbf{Threat Actor} & \cellcolor{gray!30} \textbf{0.98} & 0.78 & 0.87 & \cellcolor{gray!30} \textbf{0.98} & \cellcolor{gray!30} \textbf{0.80} & \cellcolor{gray!30} \textbf{0.88} & \cellcolor{gray!30} \textbf{0.98} & 0.77 & 0.86 \\
\textbf{Victim Country} & 0.84 & 0.77 & 0.80 & \cellcolor{gray!30} \textbf{0.88} & \cellcolor{gray!30} \textbf{0.86} & \cellcolor{gray!30} \textbf{0.86} & 0.82 & 0.77 & 0.79 \\ 
\textbf{Zero-day} & \cellcolor{gray!30} \textbf{0.96} & 0.65 & 0.77 & 0.95 & \cellcolor{gray!30} \textbf{0.95} & \cellcolor{gray!30} \textbf{0.95} & 0.89 & 0.74 & 0.81 \\
\midrule
\textbf{Average} & 0.93 & 0.73 & 0.81 & \cellcolor{gray!30} \textbf{0.94} & \cellcolor{gray!30} \textbf{0.87} & \cellcolor{gray!30} \textbf{0.90} & 0.90 & 0.76 & 0.82 \\
\bottomrule

    \end{tabular}
    }
    \label{tab:LLM scores}
\end{table}

\begin{table}[t!]
    \centering
    \caption{
        Comparison of precision (P), recall (R), 
        and F1 scores
        between a signature-based (\eg IoCParser) and
        an LLM-based approach (\eg GPT-4-Turbo).
        Note that IoCParser is capable of 
        seeking CVE, MITRE ID, and YARA rules alone.
        (*) represents the items that 
        we adopt GPT Turbo's results
        that demonstrate the best LLM performance 
        (Section~\ref{ss:rule-based}).
    }
    \resizebox{0.75\linewidth}{!}{
        \begin{tabular}{lrrrrrrrrr}
            \toprule

\textbf{Tool} & \multicolumn{3}{c}{\textbf{IoCParser~\cite{iocparser}}} & \multicolumn{3}{c}{\textbf{GPT-4-Turbo~\cite{gpt4}}} \\ 
 \textbf{Search Item} & \textbf{P} & \textbf{R} & \textbf{F1} & \textbf{P} & \textbf{R} & \textbf{F1} \\ \midrule
\textbf{CVE} & \cellcolor{gray!30} \textbf{0.98} & \cellcolor{gray!30} \textbf{0.95} & \cellcolor{gray!30} \textbf{0.97} & 0.97 & 0.84 & 0.90 \\
\textbf{MITRE ID} & 0.97 & \cellcolor{gray!30} \textbf{0.96} & \cellcolor{gray!30} \textbf{0.97} & \cellcolor{gray!30} \textbf{0.99} & 0.93 & 0.96 \\
\textbf{YARA} & \cellcolor{gray!30} \textbf{1.00} & \cellcolor{gray!30} \textbf{0.96} & \cellcolor{gray!30} \textbf{0.98} & 0.94 & 0.86 & 0.90 \\
\textbf{Attack vector*} & -- & -- & -- & 0.89 & 0.77 & 0.83 \\
\textbf{Malware*} & -- & -- & -- & 0.74 & 0.70 & 0.72 \\
\textbf{Target sector*} & -- & -- & -- & 0.82 & 0.89 & 0.85 \\
\midrule
\textbf{Average} & \textbf{0.98} & \textbf{0.96} & \textbf{0.97} & \textbf{0.89} & \textbf{0.83} & \textbf{0.86} \\ 
\bottomrule

        \end{tabular}
        }
        \label{tab:LLM_IoC}
\end{table}

\PP{LLM Prompt and Questions Design}
We design our prompt and questions 
to enhance the accuracy of responses 
during the retrieval process of an LLM.
We systematically evaluate varying question
formulations and methodological techniques 
to identify the most effective strategies.
Following the recommendations of 
Kumarasinghe \etal ~\cite{semantic_ranking},
we incorporate both the ``Role Play'' and 
``Specificity and Precision'' means for prompt generation.
The former approach enables the LLM 
to adopt a defined perspective, thereby 
generating more contextually 
appropriate and relevant responses, while
the latter approach reduces the likelihood of 
irrelevant or inaccurate outputs.
Likewise, for query construction, 
we employ ``Specificity and Precision''
to ensure simplicity and straightforwardness, 
which minimizes ambiguity and improves
overall quality of the retrieved responses.
\autoref{fig:prompt} and
\autoref{tab:questions}
in the Appendix are the final prompt 
and questions for our study.

\PP{LLM Model Choice and Evaluation}
To obtain the desirable answers
as reliably as possible, 
we sample three items 
for retrieval (\eg threat
actors, victim countries, zero-day
vulnerabilities if any).
To this end, we evaluate 
three popular LLMs: 
Gemini-1.5-Flash~\cite{geminiFlash}, 
GPT-4o~\cite{gpt4o}, and GPT-4-Turbo~\cite{gpt4}
with the same prompt and questions (\autoref{fig:prompt} and \autoref{tab:questions} in Appendix).
However, it is well-known that LLM models 
may generate 
inaccurate responses (\ie hallucinations).
This means we cannot completely 
trust the responses from an LLM.
In response, we 
randomly picked around $120$ articles
for manual inspection:
\ie a human compares the LLM responses with the 
ground truth in a technical report.
Note that we use 
precision ($P=\frac{TP}{TP+FP}$), 
recall ($R=\frac{TP}{TP+FN}$), and 
F1 ($F1=\frac{2PR}{P+R}$)
for evaluation metrics where
TP, FP, and FN denote the number of
true positives, false positives,
and false negatives, respectively.
\UP{Notably, FPs refer to cases where the 
LLM incorrectly identifies attributes 
that are not present in the report, while 
FNs arise when the model fails to detect 
attributes that are present.}
For interested readers, 
we present a selection of human-verified answers
at \autoref{tab:Appendix_manualAnswers} in Appendix.
\autoref{tab:LLM scores} presents 
a performance
comparison of three LLM models.
GPT-4-Turbo achieves
the highest scores in all Precision ($0.94$), 
Recall ($0.87$), and F1 score ($0.90$)
over other LLM models.

\subsubsection{Rule-based Retrieval}
\label{ss:rule-based}
The IoCParser~\cite{iocparser} tool
is designed for processing IoCs from 
various data sources
(\eg CTI reports, security logs, 
other security-related texts).
As the parser extracts 
three specific types of information,
including CVEs, MITRE ATT\&CK Technique IDs, 
and YARA rules, we compare it with GPT-4-Turbo.
\autoref{tab:LLM_IoC} demonstrates 
that IoCParser 
surpasses GPT-4-Turbo in extracting 
CVEs, MITRE IDs, and YARA rules,
achieving F1 scores of $0.97$, $0.97$, 
and $0.98$, respectively.
Hence, we decided to include IoCParser's
results as a complementary tool.
Our further examination of IoCParser's 
retrieval failure reveals that \WC{1} some TRs 
do not follow the standardized CVE format 
(\eg \cc{year-vul\_id} instead of 
\cc{CVE-year-vul\_id}), which reduces
the recall; \WC{2} some TRs extracted from web pages
have the parser retrieve irrelevant IoCs due to
a noise in the extracted content, 
decreasing its precision.
\begin{table}[t!]
    \centering
    \caption{
    We investigate 10 items from each technical
    report on a specific APT.
    We adopt rule-based 
    (\eg IoCParser~\cite{iocparser})
    and LLM-based approaches.
    The number (ratio) of TRs denotes 
    retrieved items out of 1,509 TRs
    because not every piece of information
    was available
    (Section~\ref{ss:availabe-data}).
    }
    \resizebox{0.4\textwidth}{!}{
        \begin{tabular}{llrr}
            \toprule

\textbf{Search Item} & \textbf{Retrieval Approach} & \textbf{\# of TRs} & \textbf{Ratio} \\ 
\midrule
\textbf{CVE} & Rule &  416 & 27.6\% \\
\textbf{MITRE ID} & Rule &  175 & 11.6\% \\
\textbf{YARA} & Rule &  131 & 8.7\% \\\hline
\textbf{Threat actor} & LLM & 1,089 & 72.2\% \\
\textbf{Victim country} & LLM &  886 & 58.7\% \\
\textbf{Zero-day} & LLM &  839 & 55.6\% \\
\textbf{Attack vector} & LLM & 1,186 & 78.6\% \\
\textbf{Malware} & LLM & 1,287 & 85.3\% \\
\textbf{Target sector} & LLM & 1,228 & 81.4\% \\
\textbf{Attack duration} & LLM, Manual inspection &  235 & 15.6\% \\ 
\bottomrule

        \end{tabular}
    }
    \label{tab:available_data}
\end{table}
As a final note, 
IoCParser does not have features
to retrieve other items.

\begin{figure*}[t!]
    \centering
    \includegraphics[width=1\linewidth, trim={20 190 20 120}, clip]{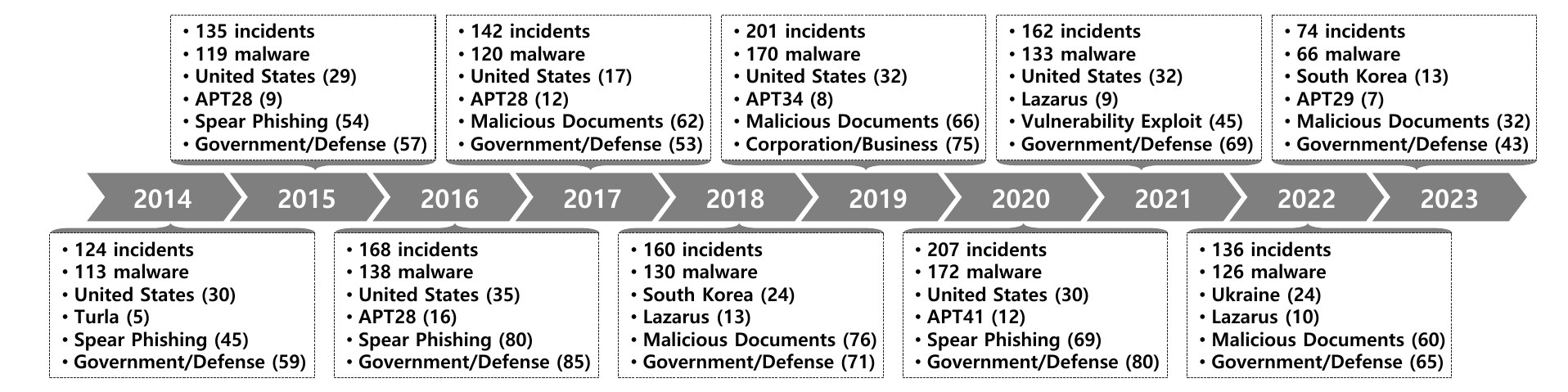}
    \caption{
    Summary of global APT trends 
    over the past decade. 
    Each box represents six key points 
    for a given year:
    the number of APT campaigns, 
    the number of associated malware samples,
    the most frequently attacked country,
    the primary threat actor, 
    the most commonly used initial attack vector, and
    the most targeted sector.
    For the last 10 years, 
    the most frequently targeted country, 
    the most active APT group, 
    the most predominant attack method, 
    and most commonly targeted sector have been
    the United States, 
    Lazarus group, malicious documents 
    (and spear phishing almost equally 
    contributed), and 
    the government/defense sector, respectively
    (Section~\ref{ss:rq1}).
    Note that the numbers in 
    parentheses represent
    the occurrences per year.
    }
    \label{fig:wholeFigure}
\end{figure*}

\subsubsection{Manual Retrieval}
\label{ss:manualDuration}
We manually verify the accuracy of attack durations 
retrieved by the LLM %
to estimate
both the lifecycle of individual APT campaigns 
and the time required to patch associated vulnerabilities.
However, determining the precise lifecycle of 
an APT incident is inherently  challenging 
due to its persistence and stealthy nature. 
Such campaigns may persist for weeks, months, or even years, 
depending on the adversary’s ultimate objective 
(\eg disrupting target operations vs.
stealing sensitive data).
Hence, we define the start and end of an APT case
based on the earliest and latest known discoveries 
of related activity, using day-level granularity.
When exact dates are unavailable, we approximate
using the midpoint of the month (\ie the 15th day).
Note that we exclude cases where the 
year-based (rough) information 
is solely available because such imprecision 
could lead to inaccurate estimations.

\subsubsection{Availability of Retrieved Information}
\label{ss:availabe-data}
As one might imagine, not every piece of information
was available in our collection of TRs. 
Indeed, we discover that only one TR 
contained all 10 items of interest.
\autoref{tab:available_data} 
presents the number of TRs
(out of $1,509$) that include 
each retrievable item.
YARA rules and MITRE ID are sparsely available,
appearing solely $8.7\%$ and $11.6\%$ of the
TRs, respectively.
In contrast, malware information and target sectors 
are prevalent, in approximately 8 out of 10 TRs 
($85.3\%$ and $81.4\%$, respectively).
Given that our analysis relies on the records in
the collected TRs as a best-effort approach, 
we find it encouraging that a non-negligible 
number of items was successfully retrieved.
We hypothesize that the covert nature of 
APT campaigns contributes to the 
limited availability of malware samples
and lack of precise information regarding
the duration of attacks.

\subsubsection{\UP{Information Retrieval from Technical Reports}}
We \UP{convert} web-based technical reports 
into PDF format using \cc{pdfkit}~\cite{pdfkit} 
in conjunction with \cc{wkhtmltopdf}~\cite{wkhtmltopdf}.
Then, we utilize LangChain~\cite{langchain} 
that offers a structured framework for data
processing and retrieval.
\UP{To enable context-aware question 
answering, we employ a 
Retrieval-Augmented Generation~\cite{RAG} 
(RAG) pipeline by vectorizing the technical 
reports into a vector database.
For each query, relevant passages are 
retrieved and incorporated into the prompt 
to provide contextually rich information to 
the LLM.
}
\UP{In summary,} our approach consists of the following phases:
\WC{1} extracting texts 
from each PDF document
using PyPDFLoader~\cite{pypdfLoader},
\WC{2} vectorizing those texts with the 
OpenAI's~\cite{openAIDocumentation} embedding model,
\WC{3} storing vectorized embeddings 
in the FAISS library~\cite{FAISS}, and
\WC{4} generating responses by
constructing prompt templates and 
chaining them with user queries to 
facilitate effective question answering.
For performance comparisons 
(\autoref{tab:LLM scores}), we access LLMs
via LangChain’s API integrations, including
GPT-4-Turbo~\cite{gpt4}, 
GPT-4o~\cite{gpt4o}, and 
gemini-1.5-flash-latest~\cite{geminiFlash}.
\UP{It is worth noting that achieving full 
accuracy and reliability remains 
challenging due to complications from PDF 
processing (Section~\ref{sec:discussion}).}

\subsection{Refining LLM-generated Responses}
\label{subsec:sanitize}

This section describes the refinement
process of LLM-generated responses
using our prompt and questions.

\PP{Normalization and Categorization}
For our analysis, it is essential to 
interpret and normalize the responses 
generated by the LLM, particularly 
for the victim countries, attack vectors, 
and target sectors. 
To ensure consistency, 
victim countries were standardized 
by converting them into 
their corresponding two-letter country codes~\cite{twoLetterCountryCode}.
For the categorization of attack vectors, 
we referred to the work of Sharma~\etal~\cite{APT_new_survey}, 
which identifies the most common attack vectors. 
This process results in 
classifying the attack vectors 
into the following 12 distinct categories: 
Spear Phishing, Phishing, Watering Hole, Credential Reuse, Social Engineering, Vulnerability Exploitation, Malicious Documents, Covert Channels, Drive-by Download, Removable Media, Website Equipping and Meta Data Monitoring.
\UP{Meanwhile, for the categorization of 
target sectors, 
we adopt the taxonomy proposed in~\cite{sectorGrouping_sectrio_apt_guide}, 
which includes nine categories: Government 
and Defense Agencies, Corporations and 
Businesses, Education and Research 
Institutions, Critical Infrastructure 
(\eg transportation, water supply), 
Financial Institutions, Individuals, 
Media and Entertainment Companies, 
Non-Governmental Organizations (NGOs) 
and Nonprofits, and Manufacturing.
In line with~\cite{sectorGrouping_google}, 
we treat Energy and Utilities and 
Healthcare as distinct categories from 
critical infrastructure. 
Additionally, we include Cloud/IoT 
Services as a category to capture its 
relevance in recent attacks, 
such as those involving supply chains.
This process results in classifying
the target sectors into 
12 distinct classes.
}

\PP{De-duplication and Filtering}
After consolidating all sources 
from the collection of TRs and TAs,
we identified a substantial number 
of duplicate entries. 
To address this, we extracted the full
text from each PDF and computed cosine similarity
using OpenAI's embedding~\cite{openai2024textembedding3small}.
We apply an empirically-chosen similarity threshold of $0.85$.
This reduces the number of TRs 
to $1,560$ unique dossiers
after removing $1,003$ duplicates.
Furthermore, we found that $51$ TRs focus on
APT trends involving multiple campaigns
rather than analyzing individual APT instances. 
We filter out these reports,
resulting in a final dataset of $1,509$ TRs.
Similarly, the number of TAs was reduced
to $884$ unique APT groups after trimming $800$ duplicates.
Next, we further filtered out the
threat actors (\textit{n} = $281$; 31.8\%)
that hold little information
beyond their names,
such as aliases and country of origin.
Finally, we use $603$ identifiable 
APT groups for further analysis.

\PP{APT Name Aliases}
The same APT group may operate under
multiple aliases.
For instance, the notorious APT group
known as APT28~\cite{APT28Descr} is 
frequently referred to as FANCY BEAR, 
Pawn Storm, or Sofacy 
depending on the security vendor.
\UP{To minimize complications, we de-duplicate APT groups with the following two-phase process.
First, we use the identifiers as primary names from each threat actor source, along with their known aliases. 
Then, we merge alias entries across different sources based on 
these primary names. 
It is possible that despite these steps, 
due to the inconsistent naming conventions 
of APT groups across vendors and the 
absence of ground truth, some inaccuracies 
may remain in our final set.
}

\subsection{\UP{Visual Representations}}
We develop the APT map 
(\autoref{fig:aptMap} in Appendix) 
and its corresponding timeline chart using 
the React framework~\cite{react} 
and the amCharts library~\cite{amchart} 
to enable interactive user experiences.
The front-end has been integrated with 
a Flask-based backend~\cite{flask}, 
deployed on the Heroku platform~\cite{heroku}.
Additionally, we use Tableau~\cite{tableau},
a visual analytics platform, to construct
a flow diagram illustrating the relationships 
between years, threat actors, and victim countries 
by incorporating the \texttt{Sankey Viz} extension~\cite{sankeyViz} 
(\autoref{fig:sankeyTop30Countries} 
in Appendix).

\section{Decadal Landscape of APT Campaigns}
\label{sec:eval}

To longitudinally understand the landscape of APTs,
we define the following 
research questions (RQs) under four themes.
\begin{itemize}[leftmargin=*]
    \item \PP{RQ1: Evolution of APTs  over a decade}
    How have APT campaigns evolved over the past 10 years regarding victim countries, 
    threat actors, target sectors, initial attack vectors, and zero-day vulnerabilities?
    \item \PP{RQ2: Cyber Threat Intelligence 
    records for APTs}
    How is APT-related information captured 
    across common threat intelligence sources, 
    including vulnerability databases, 
    attack frameworks, and indicators of 
    compromise?
    \item \PP{RQ3: Common traits of APTs}
    To what extent do APT campaigns exhibit 
    concealment and aggressiveness?
    \item \PP{RQ4: External dynamics affecting APTs}
    How do external factors, such as political events, international conflicts, global pandemics, or economic instabilities, affect APT activity?
\end{itemize}

\begin{figure*}[t!]
    \centering
    \begin{subfigure}[t]{0.47\textwidth}
        \centering
        \includegraphics[width=\linewidth]{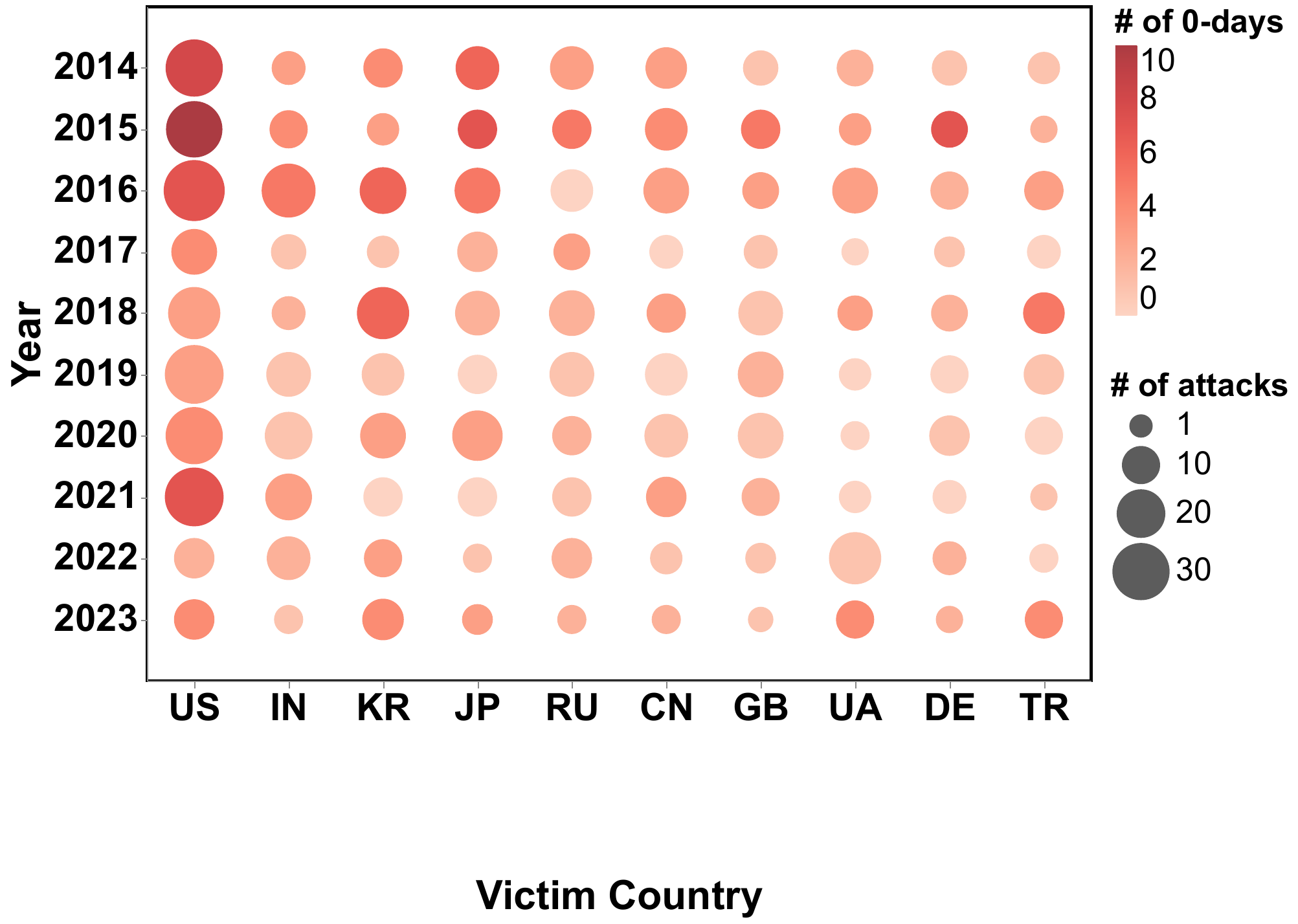}
        \caption{
        Number of APT attacks against the 
        10 most victimized countries 
        over the past decade. 
        The United States has experienced the highest
        number of attacks, followed by India, 
        South Korea, Japan, and Russia.} 
        \label{fig:victimChanges}
    \end{subfigure}
    \hfill
    \begin{subfigure}[t]{0.47\textwidth}
        \centering
        \includegraphics[width=\linewidth]{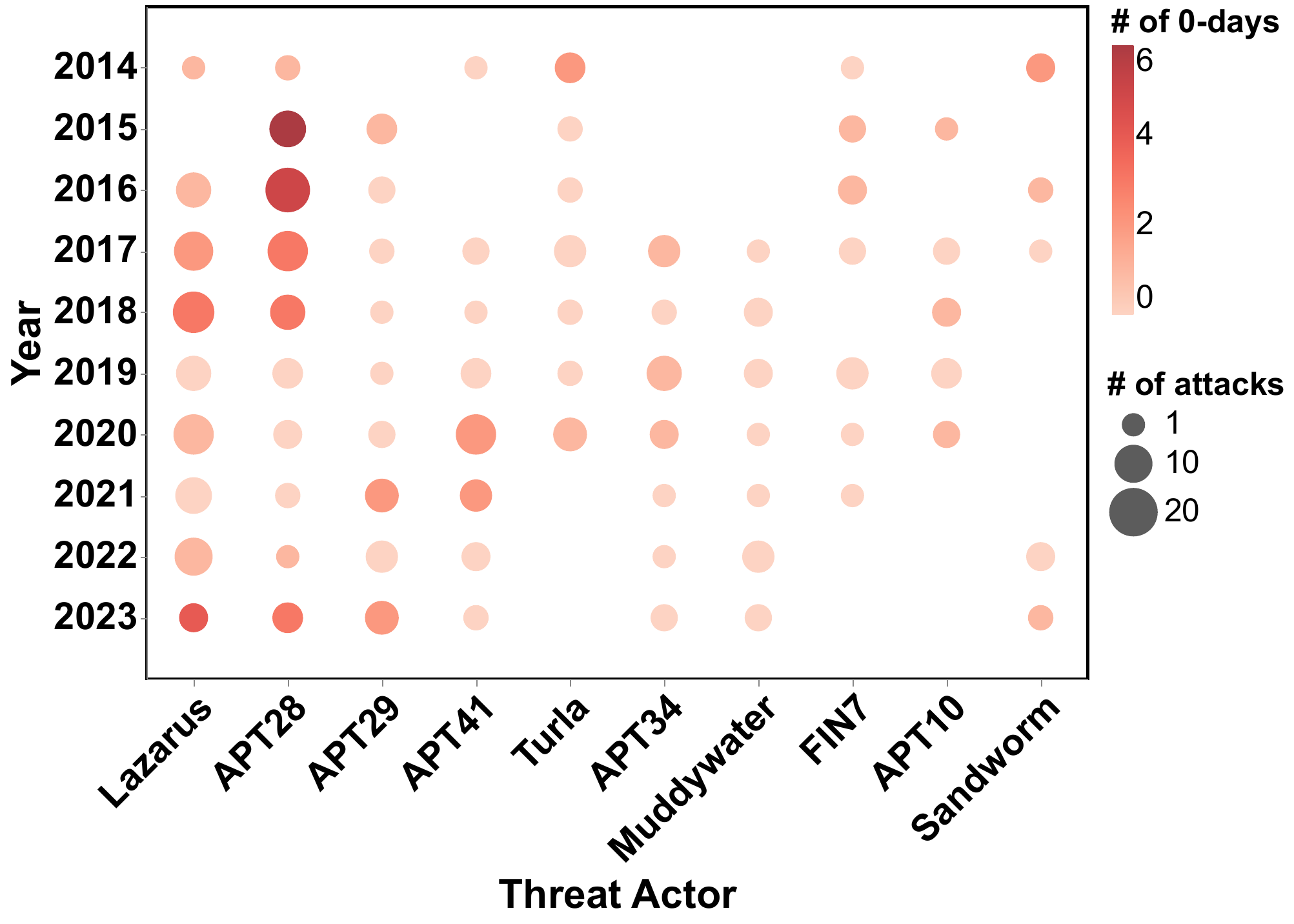}
        \caption{
        Number of APT campaigns conducted by 
        the 10 most active threat actors 
        over the past decade. 
        The Lazarus group has been 
        identified as the most active 
        APT group, while APT28 
        is notable for its frequent 
        exploitation of zero-day vulnerabilities.
        }  
        \label{fig:threatActorChanges}
    \end{subfigure}
    \caption{
    Decadal trends in APT activity 
    by victim countries (left) and 
    threat actors (right). 
    A circle size reflects the frequency of 
    APT incidents, while 
    \UP{color gradation represents the number of zero-day vulnerabilities associated with each entity 
    as a concrete value
    (\ie lighter red indicates fewer occurrences)} 
    (Section~\ref{ss:rq1}).
    }
    \label{fig:victimThreatActorChanges}
\end{figure*}

\begin{figure*}[t!]
    \centering
    \begin{subfigure}[t]{0.49\textwidth}
        \centering
        \includegraphics[width=\linewidth]{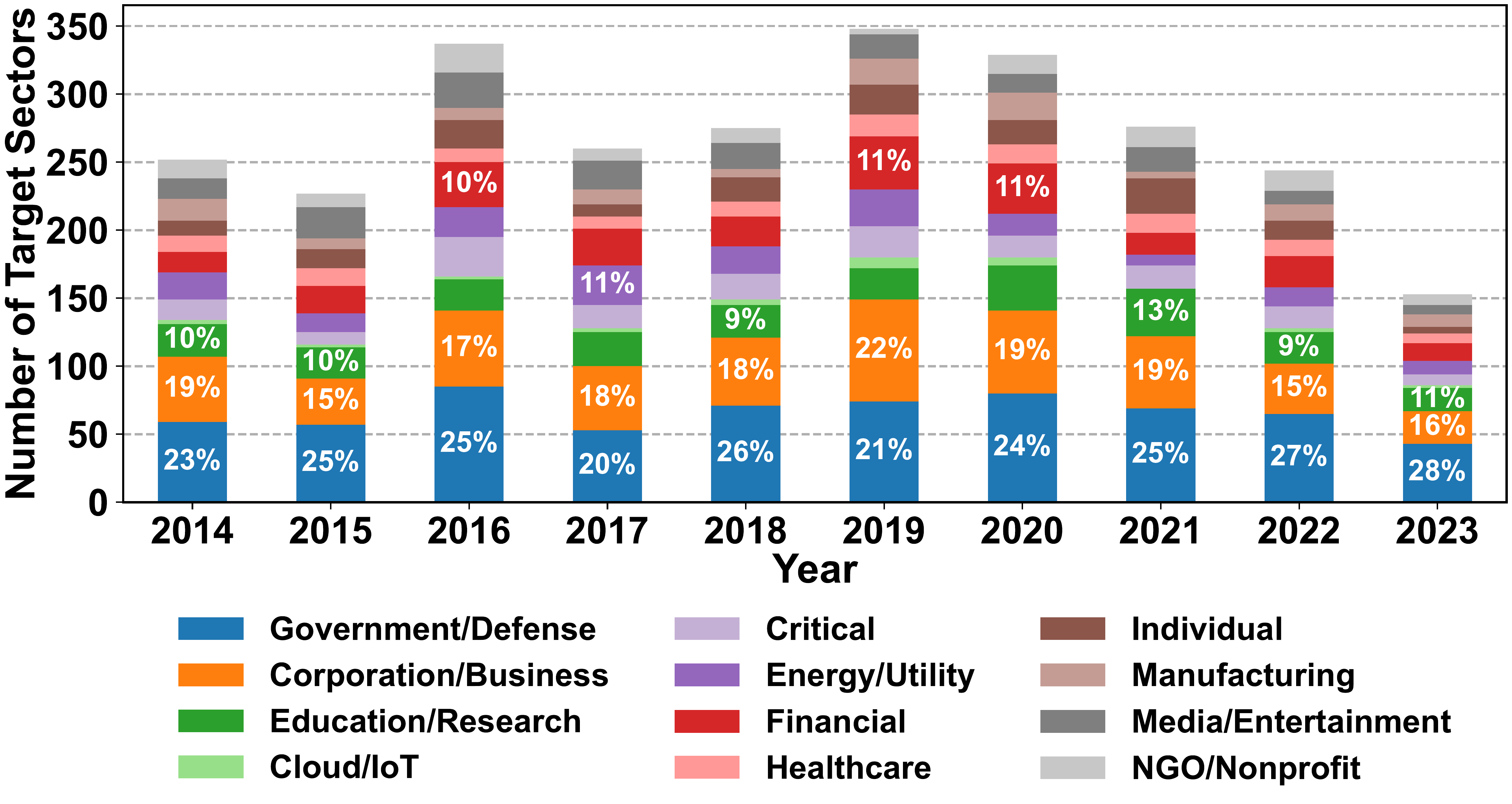}
        \caption{
        APT campaigns consistently
        target the government and business sectors,
        as well as increasing attacks 
        on the education and financial sectors
        in recent years.
        }
        \label{fig:sectorChanges}
    \end{subfigure}
    \hfill
    \begin{subfigure}[t]{0.49\textwidth}
        \centering
        \includegraphics[width=\linewidth]{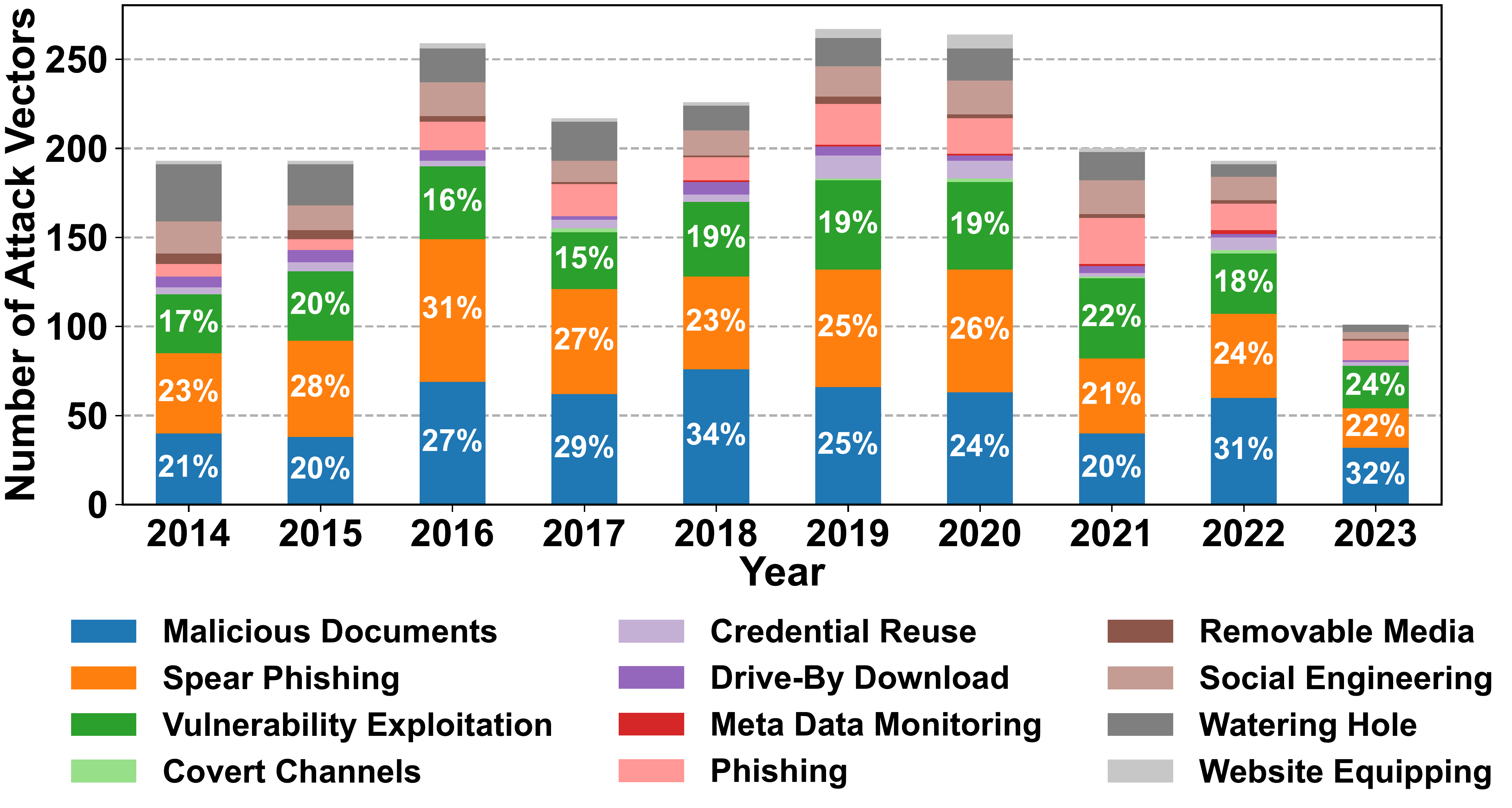}
        \caption{
        Malicious documents and spear phishing 
        remain the most common
        initial attack vectors 
        consistently harnessed by APTs, 
        followed by vulnerability exploitation.
        }
        \label{fig:attackVectorChanges}
    \end{subfigure}
    \caption{Decadal APT trends in 12 target 
    sectors (left) and 
    12 initial attack vectors (right). 
    We follow the categories of sectors from the guides~\cite{sectorGrouping_sectrio_apt_guide, sectorGrouping_google}
    and the attack vectors that Sharma \etal ~\cite{APT_new_survey} proposed.
    The figures illustrate the distributions 
    of each sector/vector over the last 10 years. 
    The percentages within a stacked bar chart
    indicate the three most common target sectors 
    and attack vectors for each year,
    along with their respective proportions. 
    Note that a single APT case may entail
    multiple target sectors or attack
    vectors, which we count individually (Section~\ref{ss:rq1}).
    }
    \label{fig:enter-label}
\end{figure*}

\subsection{Evolution of APT Campaigns Over a Decade}
\label{ss:rq1}

This section explores the temporal changes 
in APT campaigns.
\UP{We use the end year of the attack 
identified by the LLM, since the start 
dates of APT operations are often unknown 
or partially known (around 27\%) due to 
the covert nature of APT operations. 
Otherwise, we base our statistics on the 
year in which the corresponding technical 
reports were published. 
Accordingly, each APT is counted only 
once, even if it spans multiple years. 
However, we note that a single APT case may 
involve multiple target sectors or attack 
vectors, each of which is counted 
separately, as illustrated in 
Figure~\ref{fig:enter-label}.
}

\PP{Comprehensive Overview}
\autoref{fig:wholeFigure} presents a comprehensive 
overview of APT trends over the past decade, 
highlighting key attributes for each year, 
including 
the most frequently targeted countries and sectors, 
the most active threat actors, and
the most widely employed initial attack vectors.
The United States (US) consistently appears 
as the primary target, while 
the Lazarus group~\cite{LazarusDescr} 
emerges as the most active threat actor 
during this period. 
Malicious documents constitute the most popular 
initial attack vector. 
Furthermore, the frequency of APT attacks
involving malware closely aligns with 
the overall trend in APT activity, 
indicating that malware remains 
a core component of most campaigns.

\PP{Victim Countries}
A total number of 154 countries
were identified as victims across
$1,509$ APT campaigns, representing
\emph{around 80\% of all nations worldwide}.
We analyze the 10 most victimized countries that
account for $43.1\%$ ($650$) of all incidents.
\autoref{fig:victimChanges} presents 
the trends in APT attacks against these countries
from 2014 to 2023. 
US remains the primary target throughout the decade
with the exceptions of South Korea (in 2018, 2023)
and Ukraine (in 2022).
Despite minor fluctuations, our findings reveal 
that the following countries have also been 
heavily targeted:
India (IN), South Korea (KR), 
Japan (JP), Russia (RU), China (CN), Great Britain (GB), 
Ukraine (UA), Germany (DE), and Türkiye (TR).
A notable spike in attacks across the 
10 most countries occurred in 2016 with 
$174$ recorded cases.
There has been a steady decline 
in APT activity since 2021, with cases dropping from
$157$ in 2020 to $121$ in 2021, and 
to $67$ in 2023.

\PP{Threat Actors}
A total of $446$ unique threat actors 
have been identified in our APT dataset, 
$263$ of which are known to be state-sponsored
(while others unknown).
The 10 most active APT groups are 
responsible for 326 attack incidents, 
representing $21.6\%$ of all campaigns. 
\autoref{fig:threatActorChanges} shows 
the number of attacks and 
the frequency of associated 
zero-day vulnerabilities by TAs. 
Among them, Lazarus~\cite{LazarusDescr}, APT28~\cite{APT28Descr}, and APT29~\cite{APT29Descr} 
stand out as the most prolific threat actors 
over the past decade, 
with $76$, $64$, and $35$ 
campaigns, respectively. 
APT28, in particular, exhibits
a strong inclination toward 
exploiting zero-day vulnerabilities, 
with 22 instances accounting for 
$34.4\%$ of its operations.
Interestingly, the activity levels of 
these threat actors are not uniformly 
distributed over time; rather,
their campaigns tend to cluster 
within specific periods, likely reflecting 
shifting strategic objectives.
For instance, the Sandworm~\cite{SandWorm} group conducted 
the majority of its operations 
within a concentrated five-year span.
The peak year for activity among 
these top actors was 2017, during 
which $48$ attacks were recorded.
Since 2021, however, a gradual decline 
in activity has been observed, 
with the number of attacks 
falling from $47$ in 2020 to $27$ in 2021, 
and further to $26$ in 2023.

\PP{Zero-Day Vulnerabilities}
One of the common beliefs is that APT campaigns 
frequently exploit zero-day 
vulnerabilities
due to their capability to bypass defenses 
and maintain stealth. 
We identify $204$ APT incidents ($13.5\%$) 
that incorporate zero-day vulnerabilities into 
their campaigns.
\autoref{fig:victimChanges} and~\autoref{fig:threatActorChanges} 
depict the annual use of zero-day vulnerabilities 
over the past decade from the perspectives of the
top 10 victim countries and threat actors.
Our analysis reveals notable peaks in 
zero-day exploitation between 2014 and 2016, 
with an average of $39.7$ cases 
affecting victim countries 
and $7.7$ cases linked to threat actors. 
The highest recorded number of zero-day 
incidents targeting countries occurred in 2015,
reaching $50$ cases.
In contrast, 2019 saw a sharp decline, 
with only nine cases affecting victim countries 
and just one attributed to threat actors.
Since 2018, a general downward trend 
in zero-day vulnerabilities has been observed,
possibly due to the increasing prevalence
of 1-day vulnerabilities (\ie non-patched
systems) or rising costs
and the complexity of developing 
zero-day vulnerabilities.

\PP{Target Sectors}
\autoref{fig:sectorChanges} 
depicts the distribution 
of 12 sectors targeted by APT campaigns 
for 10 years, demonstrating notable shifts 
in attack patterns over time. 
The Government and Defense sector consistently 
accounts for a substantial share of APT
activities, with an average of $65.6$ incidents 
per year. 
The second most targeted is the Corporations 
and Businesses sector, which has emerged 
as a prominent focus, averaging 
$48.5$ incidents annually.
Notably, the third most frequently 
targeted sector 
varies across the years. 
The Education and Research 
sector ranks third in 6 out of 10 years, 
with a peak of $35$ cases in 2021
(possibly due to the COVID-19 pandemic). 
Meanwhile, the Critical Infrastructure sector 
has seen fluctuations,
peaking $29$ in 2016 and declining since 2019.

\PP{Initial Attack Vectors}
\autoref{fig:attackVectorChanges} illustrates 
the distribution of 12 initial attack vectors 
in APT attacks 
over the past decade. 
According to our investigation, 
malicious documents
appear the most preferred
initial access in APT campaigns, averaging 
$54.6$ incidents per year.
The second most common attack vector is 
spear phishing, 
with an average of $53.6$ cases annually.
This sophisticated phishing often entails 
sending fraudulent emails 
or deceptive messages to carefully selected 
individuals or organizations, rendering 
the attack more convincing.
Malicious documents are often combined with 
spear phishing to gain the initial access 
to target system.
Furthermore, across the whole decade, 
vulnerability exploitation consistently 
ranks as the third most frequently 
employed attack vector, with an average
of $38.9$ incidents per year.
This technique involves leveraging software 
vulnerabilities to gain unauthorized access to the system.
Interestingly, watering hole 
attacks~\footnote{A watering hole 
attack targets a specific group 
by compromising websites they frequently 
visit with malicious code. Upon successful
exploitation, attackers can gain access 
target systems and exfiltrate information while maintaining long-term control.}, 
ranked as one of most utilized 
attack vectors in 2014, 
with $32$ instances,
exhibited a gradual decline afterwards, 
with only $4$ cases recorded in 2023.

\observ{
Over the past decade, APT campaigns have impacted 
154 countries worldwide where 
US, IN, and KR remain frequently targeted. 
Although 446 unique threat actors 
have been identified, a small subset 
(Lazarus, APT28, APT29)
accounts for a significant portion of attacks.
Zero-day usage peaked between 2014 and 2016 
but has declined in recent years.
Government and corporate sectors stay 
the most targeted. 
Malicious documents and spear phishing dominate 
as initial penetration vectors}

\subsection{CTI Records in APT Campaigns}
\label{ss:rq2}
This section explores how APT campaigns can be
described across common threat 
intelligence sources, 
including a common vulnerability database, attack 
frameworks, and indicator of compromise.

\begin{table}[t!]
    \centering
    \caption{
    Top 10 most frequently observed 
    MITRE ATT\&CK techniques~\cite{MITRE_ATTaCK} 
    in APT campaigns
    over the past decade. 
    Notably, Execution, Defense Evasion, 
    and Discovery are 
    the most common tactic categories, using 
    Command and Scripting Interpreter
    and Application Layer Protocol
    being among the most prevalent.
    These tactics are well aligned with 
    the nature of APTs
    (Section~\ref{ss:rq2}).
    }
    \resizebox{0.99\linewidth}{!}{
        \begin{tabular}{lllrr}
            \toprule

\textbf{MITRE ID} & \textbf{Description} & \textbf{Tactic} & \textbf{Count} & \textbf{Ratio} \\ 
\midrule
\textbf{T1059} & Command/scripting interpreter & Execution & 77 & 3.0\% \\
\textbf{T1071} & Application layer protocol & Command and control & 76 & 2.9\% \\
\textbf{T1082} & System information discovery & Discovery & 65 & 2.5\% \\
\textbf{T1027} & Obfuscated files or information & Defense evasion & 60 & 2.3\% \\
\textbf{T1140} & \makecell[l]{Deobfuscate/decode files \\ or information} & Defense evasion & 56 & 2.2\% \\
\textbf{T1041} & Exfiltration over C2 channel & Exfiltration & 54 & 2.1\% \\
\textbf{T1204} & User execution & Execution & 51 & 2.0\% \\
\textbf{T1053} & Scheduled task/job & \makecell[l]{Execution, persistence,\\ privilege escalation} & 49 & 1.9\% \\
\textbf{T1083} & File/directory discovery & Discovery & 47 & 1.8\% \\
\textbf{T1036} & Masquerading & Defense evasion & 45 & 1.7\% \\
\bottomrule

        \end{tabular}
    }
    \label{tab:IoCs_Mitre}
\end{table}

\PP{MITRE IDs}
We extract a total of $2,582$ MITRE ATT\&CK 
techniques~\cite{MITRE_ATTaCK}
from $175$ TRs available in our collection 
(11.6\% in~\autoref{tab:available_data}),
among which $263$ technique IDs are unique.
Considering the whole $359$ distinct identifiers 
in the MITRE ATT\&CK framework,
this reveals that APT campaigns harness 
a wide spectrum of techniques 
to facilitate intrusions.
\autoref{tab:IoCs_Mitre} demonstrates 
the 10 most frequently observed techniques 
($22.5\%$ of all instances), along with 
their descriptions, tactics, occurrences, and 
proportional representation.
As expected, the most commonly observed tactics 
in APT campaigns include 
\WC{1} execution that 
involves the running code on local or remote systems; 
\WC{2} defense evasion that aims to bypass
detection mechanisms; and
\WC{3} discovery that focuses on reconnoitering 
intelligence about internal systems or networks. 
Additional frequently observed tactics are consistent
with the advanced and stealthy characteristics
of APT campaigns, such as command and control, 
exfiltration, persistence, and privilege escalation. 

\PP{CVEs}
We extract a total of $1,088$ CVEs
from $416$ TRs available in our collection 
(27.6\% in~\autoref{tab:available_data}),
among which $431$ CVEs are unique.
It is noted that multiple CVEs may be leveraged
in a single APT campaign to achieve 
full-chain exploitation.
Notably, the vulnerabilities exhibit high severity levels, 
with an average score of $8.5$.
\autoref{tab:IoCs_CVE} presents 
the top 10 most frequently observed CVEs
($22.8\%$ of all instances), 
along with their severity scores, 
vulnerability types, 
the number of affected software, 
occurrences, and proportions.
Our further analysis shows that 
Microsoft Windows and Office are 
the most commonly targeted platforms/software, 
relating to $90\%$ of the identified CVEs.
Unsurprisingly, the most prevalent 
vulnerability types include remote code 
execution (RCE), memory corruption, use-after-free (UAF),
stack overflow and privilege escalation (PE).
Notably, the CVE with the highest severity score 
($9.3$), CVE-2015-5119, is associated with 
the exploitation of Adobe Flash Player, which
was officially discontinued back in 2020.

\PP{YARA Rules}
We extract a total of $419$ YARA rules
from $131$ TRs available in our collection 
(8.7\% in~\autoref{tab:available_data}),
among which $419$ rules are unique.
We hypothesize that the limited availability 
of YARA rules may be attributed to 
the sensitive nature of APT campaigns. 
Contributing factors include: 
\WC{1} ongoing private investigations, 
\WC{2} disclosure restrictions imposed 
by non-disclosure agreements or internal policies, 
\WC{3} the risk of rule evasion or misuse 
by threat actors, and 
\WC{4} high variability and 
obfuscation in malware samples, 
which complicates the creation of generalizable 
detection rules.

\observ{
APT campaigns leverage a diverse set of 
intrusion techniques, with 2,582 MITRE ATT\&CK 
instances, highlighting frequent use of tactics 
like execution, defense evasion, and discovery.
Analysis of 1,088 CVEs reveals that 
Windows operating systems are the 
most targeted platforms, 
with remote code execution 
being the most common vulnerability type.
The CVEs are highly severe, averaging  
the score of 8.5.
On the other hand, YARA rules are scarce 
in public APT reports, 
likely due to confidentiality concerns, 
evasion risks, and the technical challenges 
posed by malware variability}

\begin{table}[t!]
    \centering
    \caption{
    Top 10 most frequently exploited 
    vulnerabilities in APT campaigns. 
    The APT groups show a strong 
    preference for 
    remote code execution (RCE) and
    memory corruption vulnerabilities,
    with CVE-2012-0158 
    being the most widely exploited (5.4\%).
    High severity vulnerabilities such as privilege escalation (PE) and
    use-after-free (UAF) 
    are also commonly leveraged.
    (Section~\ref{ss:rq2}).
    Note that we follow
    vulnerability naming
    from the National Vulnerability 
    Database~\cite{NVD}.
    }
    \resizebox{0.99\linewidth}{!}{
        \begin{tabular}{lllrrr}
            \toprule

\textbf{CVE} & \textbf{Severity} & \textbf{Vuln} &\textbf{Affected S/W} &  \textbf{Count} & \textbf{Ratio} \\
\midrule
\textbf{CVE-2012-0158} & 8.8 (High) & RCE &19& 59 & 5.4\% \\ \hline
\textbf{CVE-2017-11882} & 7.8 (High) & \makecell[l]{Memory\\ Corruption} &4& 44 & 4.0\% \\ \hline
\textbf{CVE-2017-0199} & 7.8 (High)  & RCE  & 8& 33 & 3.0\% \\ \hline
\textbf{CVE-2018-0802} & 7.8 (High)  & \makecell[l]{Memory\\ Corruption}  & 4& 20 & 1.8\% \\ \hline
\textbf{CVE-2015-5119} & 9.8 (Critical)  & UAF &  7& 18 & 1.7\% \\ \hline
\textbf{CVE-2015-1641} & 7.8 (High) & \makecell[l]{Memory\\ Corruption}  &11& 16 & 1.5\% \\ \hline
\textbf{CVE-2010-3333} & 7.8 (High) & \makecell[l]{Stack\\ Overflow} & 8& 15 & 1.4\% \\ \hline
\textbf{CVE-2014-6332} & 9.3 (High)  & RCE & 11 & 15 & 1.4\% \\ \hline
\textbf{CVE-2015-1701} & 7.8 (High)  & PE & 3& 15 & 1.4\% \\ \hline
\textbf{CVE-2014-4114} & 7.8 (High)  & RCE & 10 & 13 & 1.2\% \\
\bottomrule

        \end{tabular}
    }
    \label{tab:IoCs_CVE}
    \vspace{-15px}
\end{table}

\subsection{Common Traits of APT Campaigns}
\label{ss:rq3}
This section explores the underlying
characteristics of APTs, mostly focusing on 
concealment and aggressiveness.

\PP{APT Duration}
\begin{figure}[t!]
    \centering
    \includegraphics[width=0.75\linewidth, trim={0 0 0 0}, clip]{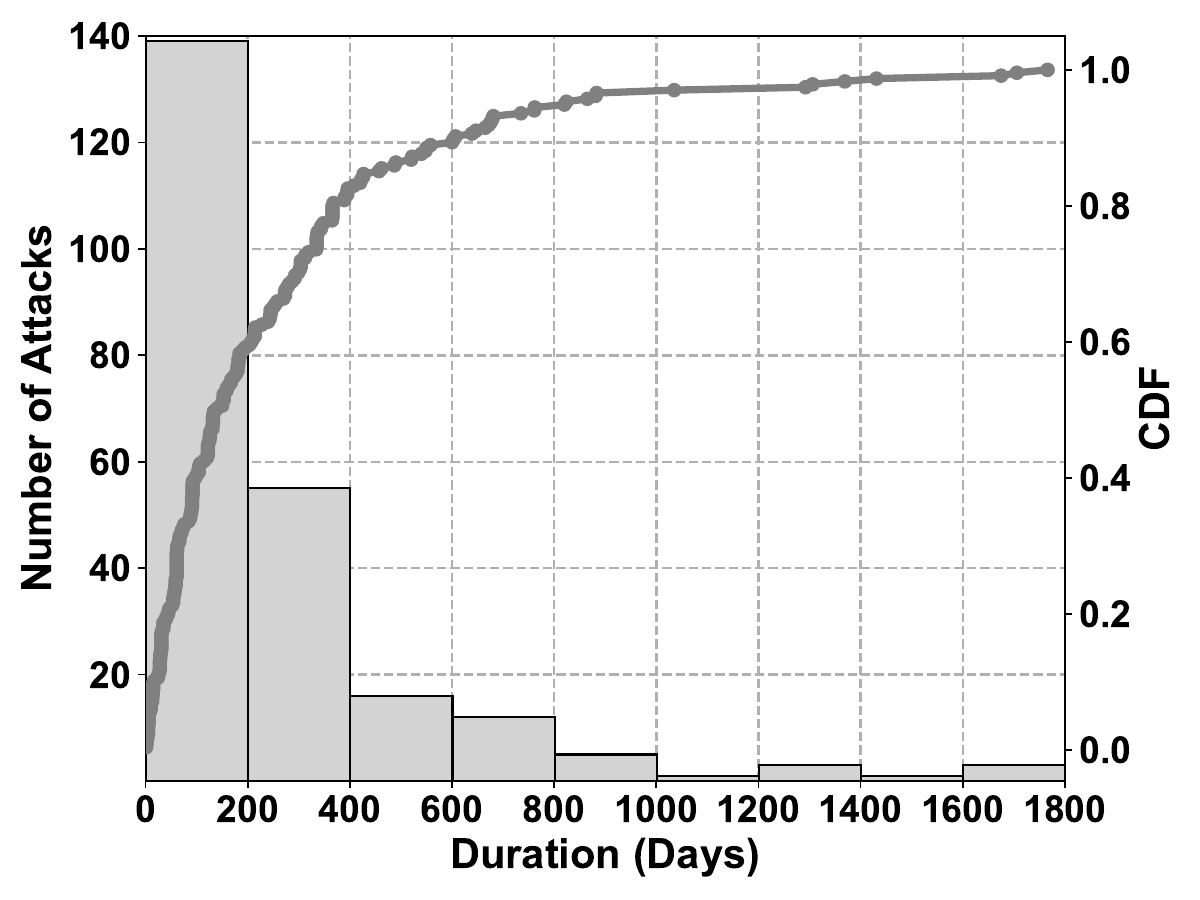}
    \caption{
    CDF and histogram of APT campaign durations
    from 235 cases (\autoref{tab:available_data}). 
    The plot shows that around half of APT 
    incidents lasted 137 days or fewer, 
    while the remaining half extended 
    beyond this duration. 
    Notably, the longest recorded campaign 
    persisted for 1,766 days, whereas 
    the shortest lasted only a single day
    (Section~\ref{ss:rq3}).
    }
    \label{fig:cdfPlot}
\end{figure}
\autoref{fig:cdfPlot} presents 
the distribution of attack durations
and their cumulative distribution function (CDF). 
Our findings indicate that approximately 
half of APT campaigns have lasted five months 
or less ($137$ days), while the remaining half 
have extended beyond that duration. 
We observe a significant variation in
attack durations, ranging from 
a single day to nearly five years.
The longest recorded campaign 
spans from June 2011 to 
April 2016 ($1,766$ days), which 
is associated with Project 
Sauron~\cite{projectSauronTR}. 
Attributed to a suspected 
US-based threat actor, this campaign
primarily targets governmental and 
research institutions in 
Russia, Iran, Rwanda, and Italy. 
The second longest campaign lasts $1,706$ days, 
and was linked to the Iranian APT group 
Ajax Security Team as part of Operation 
Saffron Rose~\cite{mandiantSaffronRoseTR}.
This operation was first detected 
on July 12, 2009, and remained active 
until March 2014.
At the other end of the spectrum, 
the shortest attack lasts only a single day,
which targeted France's TV5Monde 
broadcasting network~\cite{tv5mondeAttackTR}.
This incident, attributed to Cyber Caliphate
that is reportedly linked to the Islamic State 
of Iraq and Syria, results in 
a shutdown on April 8, 2015. 
The second shortest 
attack, spanned just two days, involves 
the distribution of malware via 
Hangul document 
files~\cite{koreaPowerPlantWiperTR}.
The campaign was observed between
December 9 and 11, 2014, primarily targeting
South Korea's power infrastructure.

\begin{figure}[t!]
    \centering
    \includegraphics[width=1\linewidth, trim={0 400 96 0}, clip]{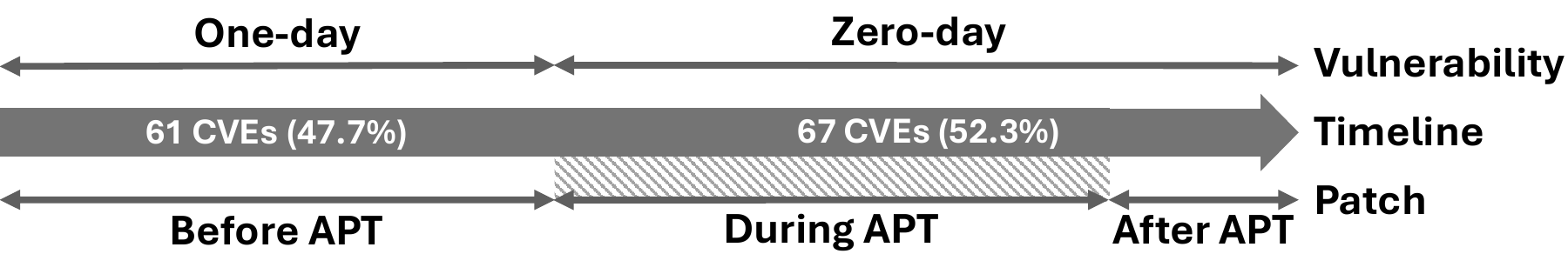}
    \caption{
    CVEs exploited in APT campaigns 
    based on patch availability 
    relative to the attack timeline. 
    We consider a zero-day vulnerability when
    CVE (with a patch) becomes available during 
    (shadowed region) or after APT campaigns.
    61 one-day vulnerabilities (47.7\%) 
    were disclosed and patched prior to 
    the APT campaign, whereas 67 zero-day 
    vulnerabilities (52.3\%) were 
    exploited before a patch was available
    (Section~\ref{ss:rq3}).
    }
    \label{fig:patching}
\end{figure}

\PP{Vulnerabilities and Patches}
We further investigate $62$ TRs
that include both CVE identifiers and 
corresponding attack durations,
recognizing $128$ distinct CVEs.
We assume that the CVE release date 
corresponds to the availability of a patch.
\autoref{fig:patching} displays the relationship
between the attack timelines and patch availability.
Our findings reveal that around 
half of those CVEs (67 or $52.3\%$) 
are exploited as zero-day vulnerabilities, 
while the remaining half CVEs (61 or 47.7\%)
had been patched prior to the associated APT campaigns.
On average, the time required to develop 
and release a patch for a zero-day vulnerability 
is approximately 200 days.

\begin{figure}[t!]
    \centering
    \includegraphics[width=1\linewidth, trim={50 140 120 180}, clip]{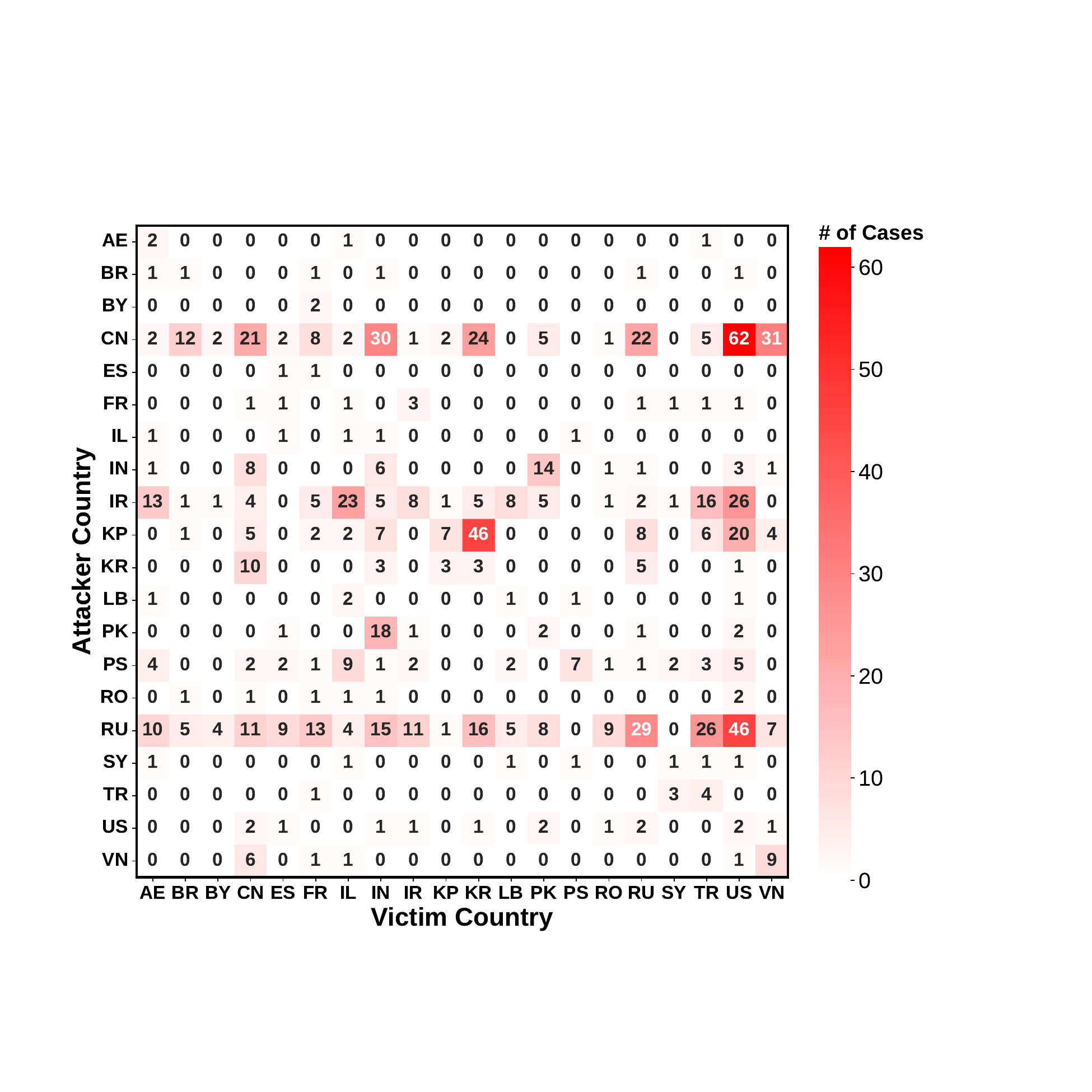}
    \caption{
    Heatmap that depicts the distribution of 
    APT campaigns based on the 20 origins 
    (attacker country) and destinations 
    (victim country), accounting for
    approximately 45\% of the whole incidents 
    over the past decade.
    The darker shades in each cell indicate 
    the higher frequencies of APT cases.
    Notable patterns include a high volume of 
    attacks originating from China, Russia, 
    and North Korea, primarily targeting 
    the United States, South Korea, 
    and other regions.
    Notably, some countries are observed to engage in self-directed attacks (Section~\ref{ss:rq3}).
    }
    \label{fig:heatMap20}
\end{figure}

\PP{Two-sided Nature as Both Attacker and Victim}
APT campaigns reveal the dual role of nations, 
where a country can simultaneously act as
both an aggressor and a victim.
The APT campaigns in our dataset reveal 
that \emph{$23$ countries have been involved 
as attackers at least once}, while 
\emph{$154$ countries have been targeted 
as victims in one or more incidents}.
We analyze the top 20 most attacking and 
victimized countries over the past decade, 
which account for $44.7\%$ of all APT cases.
Notably, \autoref{fig:heatMap20} uncovers
a significant asymmetry
between attacker and victim nations. 
For instance, the CN-US attack ratio is 31:1, 
indicating that China has conducted $31$ 
times more attacks against 
the United States than the reverse. 
Similarly, both the KP-KR and RU-US attack ratios 
stand at around 15:1, 
reflecting substantial imbalances 
in cyber offensive activity.
Russia, China, and Iran emerge
as the most active attacking country,
responsible for $229$, $232$, and 
$125$ APT cases (based on 
horizontal occurrences). 
Meanwhile, the United States, South Korea, and India 
stay the most frequently targeted,
with $174$, $95$ and $89$ incidents.
(based on vertical occurrences). 
\autoref{fig:heatMap20} also 
highlights the three most 
prominent bilateral attack relationships: namely, 
China targeting the US ($62$ cases), North Korea 
targeting South Korea ($46$ cases), and Russia 
targeting the US ($46$ cases).
It is noteworthy to mention that
not all the origin countries for
APT groups are known.

\PP{Self-directed APT Attacks}
A closer examination of~\autoref{fig:heatMap20} 
unveils that, in some APT campaigns, 
the origin and target countries are identical
(\ie main diagonal values of the matrix).
For instance, Russia has been observed 
targeting its own systems in $29$ cases, 
while China appears to have done so in $21$ cases.
Our further investigation identifies 
several contributing factors to these self-directed attacks.
First, APT campaigns may target individuals 
within the same country, often focusing on 
political dissidents or human rights 
activists~\cite{marczak_bad_2018_self_attacks, 
ovi_goldbackdoor_2023_self_attacks, 
malhotra_transparent_2021_self_attacks, 
amnesty_click_2021_self_attacks}.
These campaigns are typically conducted 
by domestic threat actors aligned with government interests.
Second, such campaigns may also be 
directed at foreign companies operating 
within the country, particularly in sectors 
such as banking and payment 
systems~\cite{google_apt32_2020_self_attacks, 
kaspersky_cloud_atlas_2019_self_attacks}.
Third, geopolitical and territorial disputes 
may drive this behavior, especially in 
politically tense regions such as the 
Russia–Ukraine border~\cite{checkpoint_cloud_2022_self_attacks, 
milenkoski_gaza_2023_self_attacks, 
grunzweig_dragonok_2017_self_attacks, 
falcone_scarlet_2016_self_attacks}.
Finally, a few self-targeted incidents 
appear to result from unusual circumstances.
For instance, the Longhorn threat actor
--linked to the United States--compromised 
a US system, only to deploy an uninstaller 
within hours, suggesting the attack was 
likely unintentional~\cite{symantec_longhorn_2017}.
Another notable case involves APT17~\cite{APT17Descr}, 
a China-associated group that reportedly 
targeted Chinese entities suspected of 
leaking sensitive information 
domestically~\cite{intrusiontruth_apt17_2019}.

\observ{
APT campaigns exhibit a wide range of durations, 
from a single day to nearly five years. 
About half of the 128 analyzed vulnerabilities 
are exploited as zero-days, while the rest are one-days.
Notably, 23 countries have acted as attackers 
and victims in APT operations, revealing 
the dual-role nature of many nations 
and significant asymmetries
(\eg China launching 31 times more attacks 
on the United States than the opposite).
Some campaigns were self-directed, 
with countries like Russia and China 
targeting domestic entities, often due to 
political repression or foreign company 
surveillance}

\subsection{External Dynamics of APT Campaigns}
\label{ss:rq4}

Although APT operations are typically goal-driven 
and backed by sponsors, uncovering behind these attacks 
or external factors inherently remains a significant challenge.
This section examines our curated collection of news articles
and media reports from the web, discussing any known
relationships (\eg triggers, motives, or impacts) 
between global affairs and APT activities.
We classify such external dynamics into four categories:
political events, international conflicts, global pandemics,
and economic gains.

\PP{Political Events}
APT groups frequently 
exploit politically sensitive 
events to further their objectives.
A notable example was 
the Russia-affiliated 
group Fancy Bear (APT28)~\cite{APT28Descr}, 
which launched
a spear-phishing campaign targeting 
the Democratic National Committee 
(DNC) during the 2016 US presidential 
election~\cite{hilaryElections2016}. 
As part of this operation, 
APT28 successfully 
breached systems linked to the 
presidential campaign, resulting in 
the theft of sensitive data.
In a similar vein, Fancy Bear targeted
Macron’s presidential campaign during 
the 2017 French election.
The group infiltrated campaign systems,
stealing credentials and 
sensitive data~\cite{MacronUS}.
Fancy Bear is also suspected of
compromising the infrastructure
of a German political party.
By infiltrating the network, the
group harvested critical information 
for political advantage~\cite{DigitalAttackGerman}.

\PP{International Conflicts}
APTs linked to international conflicts 
often involve 
the intersection of geolocational tensions and cyber warfare. 
For instance, during the Russo-Ukrainian War~\cite{rusUkrWar},
the Russian-sponsored APT group known as Sandworm~\cite{SandWorm}
hacked Ukrainian energy 
infrastructure~\cite{sandwormInternationalConflicts} back in 2015, causing
power outages that affected 230K consumers.
Sandworm also attempted to disrupt
the news agency in Ukraine~\cite{RussiaAttackUkraine}
afterwards.
Similarly, around 2023, the notorious
Russian cyberespionage group,
named Fancy Bear,
launched an attack on 
a critical energy 
facility~\cite{apt28InternationalConflicts}.

\PP{Global Pandemics}
Beginning in late 2019, the COVID-19 
pandemic brought the unprecedented 
impacts across various sectors 
worldwide~\cite{covidWiki}.
\autoref{fig:sectorChanges} 
shows a noticeable rise in
APT attacks targeting the healthcare 
sector, with $14$ cases ($4.3\%$), 
and the education and research sector, 
with $33$ cases ($10.0\%$) in 2020,
coinciding with the 
global 
spread of the virus.
During this period, the China-sponsored
APT41~\cite{APT41Descr} group exploited vulnerabilities 
in remote desktop services to attack
healthcare 
organizations~\cite{apt41GlobalCrises}. 
Similarly, the Lazarus~\cite{LazarusDescr} group
attempted to steal 
COVID-19-relevant 
intelligence~\cite{lazarusGlobalCrises}
by targeting a pharmaceutical company.
Additionally, CozyBear (APT29)~\cite{APT29Descr} 
has been suspected of
attempting to steal the COVID-19 
vaccines information~\cite{pfizerCovid2020}. 
The World Health Organization (WHO) was
not spared, as the APT group
called 
DarkHotel launched a
password-stealing attack 
against WHO staff~\cite{WHODarkHotel}.

\PP{Economic Gains}
APT activities are associated 
with the financial sector, often driven
by economic gain. 
For instance, the Carbanak group (FIN7)~\cite{FIN7}
gained notoriety for its 
cyberattacks on 
Russian banking institutions in 
2016~\cite{carbanakBanks2016}, 
coinciding with Russia's gradual 
recovery from a prolonged economic
recession~\cite{russianEconomy2016}. 
With the rise of cryptocurrencies, 
many threat actors have
shifted their focus to these 
digital assets as primary targets. 
Notably, the Lazarus group,
reportedly 
sponsored by North Korea,
was implicated to a major 
cryptocurrency theft 
in 2022~\cite{lazarusEconomic}. 
Both FIN7 and Lazarus are recognized 
for their persistent
targeting of financial institutions,
underscoring the significant economic 
motivations behind their operations.

\observ{
APT groups frequently exploit politically 
sensitive events, such as presidential 
elections, to infiltrate voting systems 
and influence public perception.
Geopolitical conflicts often coincide 
with cyber operations, with Russian
APTs launching attacks 
during international tensions.
During the COVID-19 pandemic, 
healthcare and research sectors 
experienced a spike in APT campaigns
(\eg APT41, Lazarus, APT29), 
targeting pandemic-related organizations.
Economic motives also drive APT activity
(\eg FIN7, Lazarus), 
conducting cyberattacks 
on financial institutions and 
cryptocurrency platforms
reflecting a persistent focus on financial gain}

\subsection{Visual Representations}
\label{sec:visualize}
We design an interactive map 
to visualize APT campaigns,
enabling users to explore 
detailed information by selecting 
either an attacking or victimized country.
The map integrates decade-long 
historical data by year, presenting 
key attributes such as associated 
threat actors, source(s), CVE identifier(s), 
initial attack vector(s), related malware, 
targeted sector(s), and 
estimated attack duration
(when such information is available
in the corresponding technical report).
More importantly, the map 
maintains \emph{up-to-date information 
by dynamically retrieving content} 
from technical reports using an LLM.
Note that the source has been 
currently linked to TR\#1~\cite{TR1}.
In addition, we incorporate a timeline
chart that links APT campaigns to relevant news articles for contextual reference.
\autoref{fig:aptMap} in Appendix 
presents a sample view 
of an APT campaign from 2020, in which APT41~\cite{APT41Descr}
targeted Canada.
Finally, we display an additional 
interactive diagram 
(\autoref{fig:sankeyTop30Countries} in Appendix) 
that visualizes the relationships between 
the top 10 threat actors and 
the 30 most frequently targeted countries 
over the past decade.

\section{Discussion and Limitations}
\label{sec:discussion}

This section discusses the limitations of our work.

\PP{Representativeness of APT Campaigns}
Although our dataset covers 
a wide spectrum of APT campaigns,
\UP{recent} incidents likely remain 
unreported or undocumented.
For instance, publicly available technical reports in recent 
years appear less comprehensive, as shown by the decline observed in 2023
(\autoref{tab:articles_stats}).
Besides, due to the covert nature of APT
operations, capturing every case 
is inherently infeasible.
\UP{While we chose three 
independent repositories
to include as many APT 
instances as possible, 
community-aggregated datasets
may still introduce selection biases and coverage gaps.}
Nonetheless, we believe that
the collective intelligence 
\UP{drawn from} both official and 
unofficial sources worldwide 
\UP{provides} a representative sample,
which would be sufficient to 
approximate a meaningful reflection 
of the broader ground truth.

\PP{Limited Responses from an LLM}
While extracting targeted information using 
an LLM can be highly effective,
it comes with several limitations related to 
accuracy, reliability, and completeness.
First, PDF files are unstructured, leading 
to extract misordered or misaligned content 
during extraction.
Complex layouts including tables and figures
could further result in broken or fragmented text.
Second, long documents must be segmented
into smaller chunks due to the limited 
context window of LLMs, which may
disrupt contextual coherence.
Third, LLMs are susceptible to hallucination,
potentially creating inaccurate or fabricated
information when encountering 
incomplete or ambiguous input.
Lastly, extracted text may contain extraneous or irrelevant content
such as page numbers, headers, footers, legal disclaimers, or noises.
With the possibly above reasons, 
our empirical assessments with sampled 
TRs (\autoref{tab:LLM scores})
demonstrate that the GPT-4-Turbo model
achieves around $90\%$ F1 scores.
Nevertheless, we acknowledge that 
our study represents a best-effort approach, 
as the retrieval process is inherently dependent 
on the interpretive capabilities of the LLM.

\PP{Attack Duration}
We understand that estimating the 
attack duration of APT campaigns 
is inherently challenging due to the covert and
persistent nature of these operations.
First, the reports may have inconsistent timelines
across multiple reports by different organizations.
Second, the reported dates may be distorted
due to the unwillingness of a victim or
national security reasons.
Third, reliance on public sources 
may miss internal or classified timelines.
Lastly, inferring the end date may be
open because detecting the last 
known activity is difficult to determine.
This work attempts to reconstruct durations
based on the (known) records.

\PP{CVE and Patch Timing}
In many cases, a patch is available 
at the time of CVE publication.
However, the timing of a CVE’s 
release does not always guarantee 
patch availability, as it relies on 
responsible disclosure practices. 
Note that our study aims to approximate 
the timeline required to develop 
patches for zero-day vulnerabilities 
exploited in APT campaigns.

\PP{Future Work}
\UP{While we carefully examine attack 
durations (Section~\ref{ss:rq3}) and 
attacker motivations 
(Section~\ref{ss:rq4}), there remain 
additional opportunities to gain deeper 
insights. One promising direction is to 
further investigate the relationship 
between attack duration and attacker 
objectives. Another is to uncover the 
evolution of persistent techniques and 
remediation trends over time.}

\section{Related Work}
\label{sec:relwork}

We categorize prior APT works
into three main areas, including
the detection and evaluation, Cyber
Threat Intelligence,
dataset regarding APT campaigns.

\PP{APT Detection and Evaluation}
A substantial body of research has focused 
on the detection and evaluation of APT campaigns, 
including the development of frameworks 
such as HOLMES~\cite{holmes_apt}, 
ProvG-Searcher~\cite{ProvG}, 
Zimba~\etal~\cite{ModelnDetection}, 
Marchetti~\etal~\cite{highVolumeNetwork},
and Hassan~\etal~\cite{TacticalProvenance}.
Notably, MAGIC~\cite{magic_apt} leverages 
graph neural networks and self-supervised learning 
to construct behavior graphs from system logs, 
applying a masked graph learning strategy 
to capture relationships between entities and events.
On the other hand, CAPTAIN~\cite{GradientRules} introduces 
a rule-based intrusion detection system, which
enhances traditional provenance-based approaches 
by learning fine-grained detection rules through
gradient descent optimization.
Wang~\etal ~\cite{Wang_2024} propose 
a provenance graph-based detection framework 
by reconstructing attack chains 
and countering adversarial strategies 
through empirical evaluation.
Meanwhile, Shen~\etal~\cite{mitre_edr_performance}
further contribute to the field by 
systematically evaluating  
the effectiveness of modern Endpoint Detection and 
Response (EDR) systems in detecting and mitigating 
adversarial tactics commonly used by APT groups 
in realistic attack scenarios.
Malik~\etal~\cite{Malik2024Advanced} present
a multi-layered mitigation framework, 
evaluating network-based, 
host-based, and AI-driven detection methods. 
Our work differs from previous approaches,
with a focus on trends and insights 
over the last decade from a macroscopic perspective
(\eg temporal and global analysis).

\PP{Cyber Threat Intelligence}
Several studies have concentrated on 
the collection and evaluation of CTI. 
With an in-depth analysis of 22 APT reports, 
Ussath~\etal~\cite{apt_behind_scene} identify 
common tactics, tools across different 
stages of APT campaigns. 
Similarly, Bahrami~\etal~\cite{cyber-kill} propose 
a taxonomy based on the cyber kill chain model 
to systematically categorize the tactics, techniques, 
and procedures (\ie TTPs)  by threat actors. 
Kumarasinghe~\etal~\cite{semantic_ranking} 
present a multi-stage ranking framework 
designed to identify and prioritize 
the most relevant MITRE ATT\&CK techniques 
by leveraging a combination of pre-trained and 
fine-tuned language models.
Meanwhile, TTPHunter~\cite{ttpHunter} 
introduce an automated system that 
applies machine-learning-based sentence classification 
to extract MITRE ATT\&CK-aligned TTPs from 
unstructured APT reports.
Note that our work uses an LLM to retrieve
pre-defined questions from TRs.

\PP{APT Dataset}
TREC~\cite{trec} collects a dataset by capturing APT behaviors at the kernel level, making it publicly available.
Similarly, Kumarasinghe \etal~\cite{semantic_ranking} introduce 
an open benchmark dataset for training 
and validating 
varying techniques used in APT campaigns, which
contains threat behavior descriptions from 
real-world reports and MITRE ATT\&CK technique identifiers.
Siracusano~\etal~\cite{aCTIon} 
provide a collection of publicly available 
technical reports and corresponding CTI information.
Moreover, Stojanović~\etal~\cite{APT_datasets} provide 
a comprehensive review of existing APT datasets 
and the frameworks employed to model attacks 
for the development of automated detection techniques. 
Their analysis highlights the limited availability 
of publicly accessible datasets and 
discusses the considerable challenges 
involved in collecting realistic, 
high-quality attack data. 
In this work, 
we compile several (large) collections of 
prior technical reports 
(\eg 2,563 before dataset refinement),
threat actor profiles,
and related news articles.

\section{Conclusion}
\label{sec:conclusion}
This study presents a comprehensive, 
decade-long (from 2014 to 2023) analysis 
of APT campaigns, offering a macroscopic 
perspective  of how these threats have evolved 
across countries, sectors, and 
attack techniques. 
By leveraging a hybrid information retrieval 
approach that combines LLM inference 
and rule-based extraction, 
we systematically process and analyze 
over 1,500 technical reports, 
revealing key trends in threat actor behavior, 
vulnerability exploitation, and campaign duration. 
Our key findings highlight the global reach, 
persistent nature, and strategic targeting 
patterns of APT groups, 
while also uncovering the contextual influence 
of geopolitical events, global crises, 
and economic motivations. 
Although most observations align with our
expectations, our findings also discover 
interesting (but less known) facts such as
a wide range of attack durations,
self-directed attacks, notable prevalence 
of one-day CVEs.
To support ongoing research and 
awareness in the area of APTs, 
we release an interactive visualization platform,
which promotes deeper 
engagement with the evolving threat landscape.

\section*{Acknowledgments}
We thank the anonymous reviewers 
for their constructive feedback.
This work was partially 
supported by the grants from
Institute of Information \& communications 
Technology Planning \& Evaluation (IITP),
funded by the Korean government 
(MSIT; Ministry of Science and ICT):
No. RS-2022-II221199,
No. RS-2024-00437306, 
No. RS-2024-00337414,
No. RS-2025-25457342,
and No. RS-2025-25394739.
Additional support was provided by the National Science Foundation under awards CNS-2126654, CNS-2440819, and DGE-2335798.
Any opinions, findings, and conclusions or 
recommendations expressed in
this material are those of the authors and 
do not necessarily reflect
the views of the sponsor.

\bibliographystyle{ACM-Reference-Format}
\bibliography{ref}

\appendix
\section{Appendix}
\label{sec:set-diff-dodis}

\subsection{Notable APT Groups}
Among the threat actors,
we examine the 10 most active
APT groups over the past 10 years.
The Equation Group~\cite{EquationGroup} is a highly sophisticated threat actor suspected of being tied to the USA. 
APT28~\cite{APT28Descr}, affiliated with Russian intelligence, is known for its primary targets in the government, military, and security sectors.
Lazarus Group~\cite{lazarus_SonyHack} is allegedly run by the government of North Korea. They have been responsible for several cyberattacks since 2010.
APT41~\cite{APT41Descr}, believed to operate under Chinese state-sponsored initiatives, is notorious for its dual-purpose operations, conducting both espionage and financially motivated campaigns.
Turla~\cite{Turla}, known as the Russian-linked cyber-espionage group, is recognized for its stealthy and sophisticated operations targeting government and diplomatic entities.
APT34~\cite{APT34}, suspected to be associated with Iran, focuses on cyber-espionage against government, financial, and energy sectors.
APT29~\cite{APT29Vaccine}, also linked to Russia, targeted Western governments and scientific institutions, particularly during the COVID-19 pandemic, in attempts to obtain vaccine-related data.
FIN7~\cite{FIN7}, a financially motivated group with ties to Russia, has conducted large-scale cyber crimes against the hospitality and retail sectors, stealing massive amounts of payment card data.
The MuddyWater group~\cite{Muddywater}, suspected of being linked to Iran’s government, has carried out cyber-espionage campaigns targeting government and private organizations across the Middle East, Europe, and North America.
Sandworm~\cite{SandWorm}, connected to Russian intelligence, is infamous for its disruptive attacks on critical infrastructure, including power grids and industrial control systems.
Finally, APT10~\cite{APT10} is a Chinese state-sponsored cyber-attack group, mainly targeting aerospace, engineering, and telecom firms.

\begin{table}[h!]
    \centering
    \caption{Questions to obtain
    the responses from LLM.}
    \resizebox{0.95\linewidth}{!}{
    \begin{tabular}{ll}
        \toprule
        
\textbf{Item} & \textbf{Query} \\ \midrule
\textbf{Threat Actor} & \makecell[l]{``What is the name of a threat actor group?''} \\ \hline
\textbf{Victim country} & \makecell[l]{``Which countries are being targeted?''} \\ \hline
\textbf{Zero-day} & \makecell[l]{``Was a zero-day vulnerability used in\\ this attack? Answer with TRUE or FALSE.''} \\ \hline
\textbf{Attack vector} & \makecell[l]{``What are the initial attack vectors described\\ in this report? Group them into one of followings:\\ Spear Phishing, Phishing, Watering Hole,\\ Credential Reuse, Social Engineering,\\ Exploit Vulnerability, Malicious Documents, \\Covert Channels, Drive-by Download,\\ Removable Media, Website Equipping,\\ Meta Data Monitoring.''} \\ \hline
\textbf{Malware} & \makecell[l]{``Which specific malware, tool names,\\ of software frameworks are used\\ in the attack from this report?''} \\ \hline
\textbf{Attack duration} & \makecell[l]{``Analyze the following dates extracted from\\ an article discussing an APT attack. Identify the\\ start and end dates of the attack timeline,\\ considering the dates provided may indicate\\ various stages of the attack. Return the start\\ date as the earliest reference to attack activity\\ and the end date as the latest reference of activity\\ specific to the attack. All Dates from \\article:  [ALL\_DATES\_FROM\_ARTICLE].\\Additionally, the article's publication date\\ is provided for context: [PUBLISH\_DATE].''} \\ \hline
\textbf{Target sector} & \makecell[l]{``Identify the targeted sectors in this document. \\Group them into one of followings: Government\\ and defense agencies, Corporations and Businesses,\\ Financial institutions, Healthcare, Energy and utilities, \\Cloud/IoT services, Manufacturing, Education and\\ research institutions, Media and entertainment\\ companies, Critical infrastructure, Non-Governmental\\ Organizations (NGOs) and Nonprofits, Individuals.''} \\ 
\bottomrule

    \end{tabular}
    }
    \label{tab:questions}
\end{table}

\begin{table}[t!]
    \centering
    \resizebox{0.95\linewidth}{!}{
    \begin{tabular}{llll}
        \toprule
        
\textbf{Label} & \textbf{Initial Attack Vector} & \textbf{Label} & \textbf{Target Sector}\\ 
\midrule
V1 & Spear-Phishing & S1 & Government and Defense\\
V2 & Phishing & S2 & Corporations and Businesses\\
V3 & Watering Hole & S3 & Critical\\
V4 & Credential Reuse & S4 & Financial\\
V5 & Social Engineering & S5 & Healthcare\\
V6 & Vulnerability Exploitation & S6 & Energy and utilities\\
V7 & Malicious Documents & S7 & Individuals \\
V8 & Covert Channels & S8 & Education and Research\\
V9 & Drive-by Download & S9 & Media and Entertainment\\
V10 & Removable Media & S10 & NGOs and Nonprofits\\
V11 & Website Equipping & S11 & Manufacturing\\
V12 & Meta Data Monitoring & S12 & Cloud and IoT services\\
\bottomrule

    \end{tabular}
    }
    \caption{
    Labels for initial attack vectors and target sectors. 
    }
    \label{tab:AcronymTable}
\end{table}

\begin{minipage}{1\linewidth}
    \resizebox{0.95\linewidth}{!}{

\begin{boxI}
You are an experienced security engineer
analyzing the security articles describing 
cases of APT attacks. You need to answer 
the questions in the scope of the provided 
file as precise, accurate and short as possible.

For each question just give me the answers straight 
without additional explanation of it. If the information 
is not mentioned answer with ``Not mentioned''. 
Given below are the contents of the file and 
question of the user. \\
\{"context"\} = [YOUR\_CONTEXT] \\
\{"question"\} = [YOUR\_QUESTION]
\end{boxI}

    }
    \captionof{figure}{LLM prompt to retrieve
    desirable information.}
    \label{fig:prompt}
\end{minipage}

\clearpage

\begin{figure*}[t!]
    \centering
    \resizebox{0.95\linewidth}{!}{
        \includegraphics[angle=0]{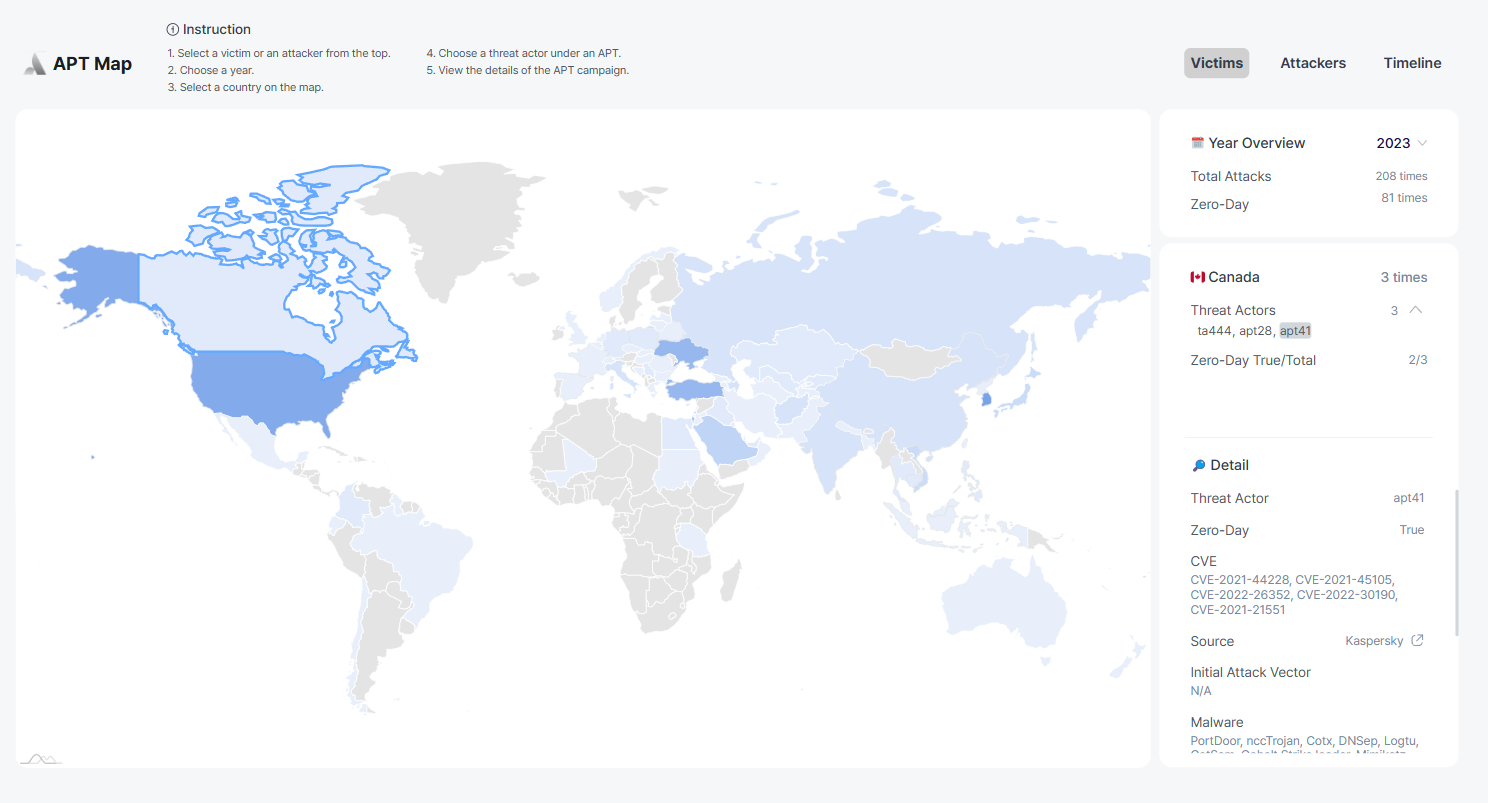}
    }
    \vspace{-0.3cm}
    \caption{Snapshot of our APT map with victim countries around 2023. 
    A total of 208 attacks and 81 involving 
    zero-day vulnerabilities were identified 
    that year. 
    Selecting a country (\eg Canada) in the map 
    reveals associated threat actors, and 
    choosing a specific actor 
    (\eg APT41~\cite{APT41Descr}) 
    displays the details of the corresponding 
    APT campaign below such as
    CVE information, malware, zero-day
    vulnerability, and initial attack vector
    (Section~\ref{sec:visualize}).
    }
    \label{fig:aptMap}
\end{figure*}

\begin{figure*}
    \centering
    \includegraphics[width=0.95\linewidth]{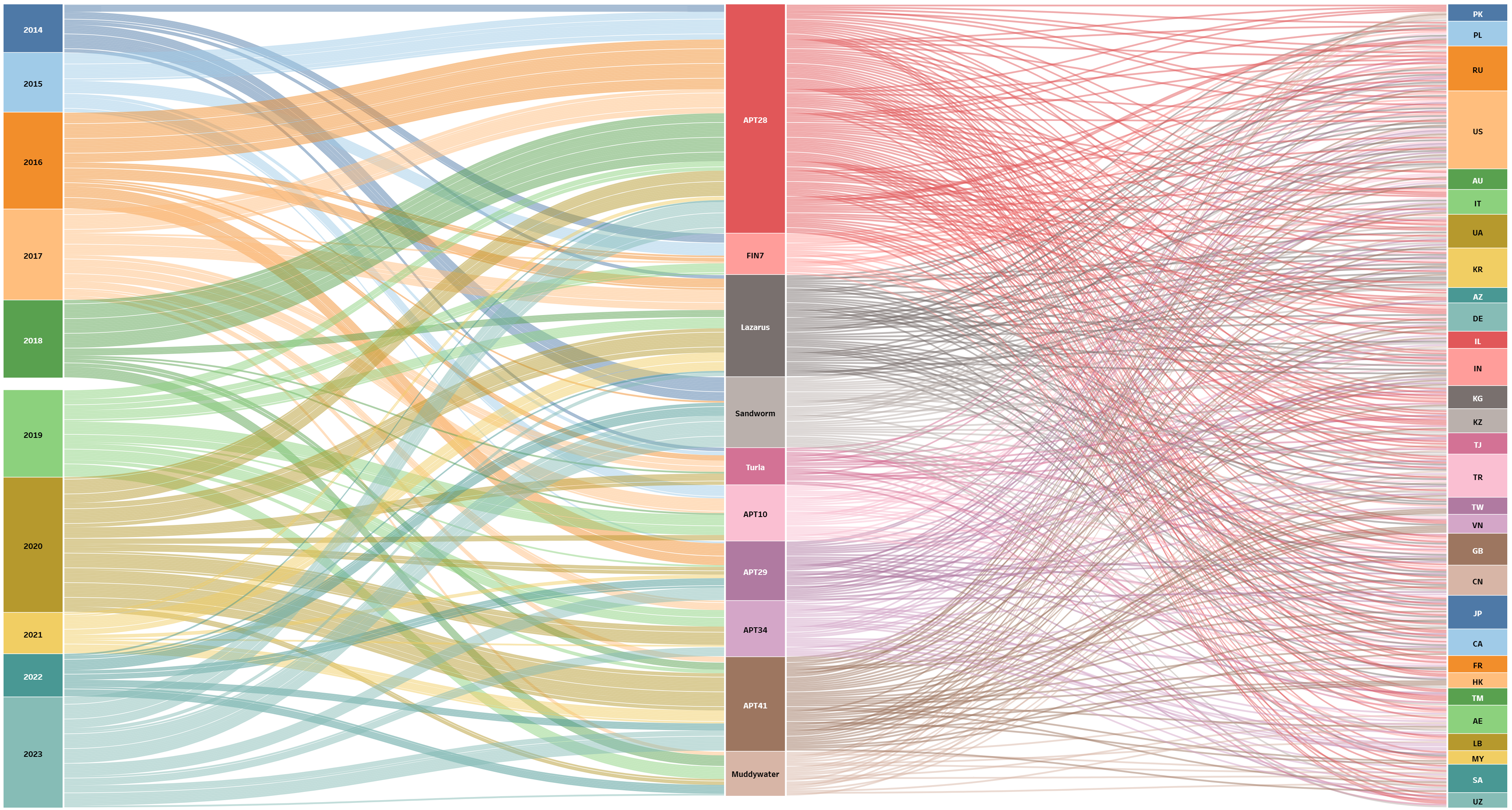}
    \caption{Flow diagram that depicts the relationships between the top 10 threat actors and the top 30 victimized countries over the past decade. 
    Three columns represent the year, threat actor and victim country, respectively  
    (Section~\ref{sec:visualize}).}
    \label{fig:sankeyTop30Countries}
\end{figure*}

\begin{table*}[t!]
    \centering
    \caption{
    Selected APT campaigns 
    along with corresponding information, 
    including a threat actor(s),
    victim country(ies), zero day (ZD) vulnerabilities, associated
    malware, attack duration, initial attack vector(s), and target sector(s).
    The curated set represents the samples used
    for manual inspection to evaluate the accuracy
    of LLM's repsonses with retrieval.
    The  complete list of label definitions 
    (\eg V1-12, S1-12) is 
    provided in~\autoref{tab:AcronymTable}.
    Note that a dash (-) indicates missing or 
    unavailable information.
    }
    \resizebox{1\linewidth}{!}{
    \begin{tabular}{llllllllllllllllllllllllllll}
        \toprule

\textbf{Year} & \textbf{Threat Actor} & \textbf{Victim Countries} & \textbf{ZD} & \textbf{Malware or Tools} & \textbf{Attack Duration} & \textbf{Attack Vectors} & \textbf{Target Sectors} \\
\midrule

2014 & Pawn Storm & \makecell[l]{US, HU, AT, DE, FR, IQ, PL, PK} & \xmark & SEDNIT & -- & V1, V2 & S1, S2, S9 \\ \midrule

2014 & Quedagh & UA, GE & -- & BlackEnergy & Apr '13 – May '14 & V5, V7 & S1 \\ \midrule

2014 & Deep Panda & US, JP, MY, HK & -- & \makecell[l]{Derusbi trojan, implants and droppers,\\ RATs, DLL backdoor} & -- & V1, V5 & \makecell[l]{S1, S2, S3, S6, S10} \\ \midrule

2014 & NEWSCASTER & \makecell[l]{US, IL, UK, SA, SY, IQ, AF} & \xmark & IRC Bots, data profiling tools & -- & V2, V5 & \makecell[l]{S1, S2, S7} \\ \midrule

2015 & Naikon APT & \makecell[l]{PH, MY, KH, ID, VN, \\ MM, SG, TH, NP, LA} & \xmark & \makecell[l]{Remote administration utility, keyloggers,\\ WinPcap network sniffers, RTLO, spyware} & -- & V1 & S1, S2 \\ \midrule

2015 & Pawn Storm & US, CA & \cmark & Keyloggers, C\&C servers, sandbox & -- & V1 & \makecell[l]{S1, S2, S7} \\ \midrule

2015 & Butterfly & \makecell[l]{BR, CN, HK, IN, IL, JP, KZ, \\ MY, MA, NG, TW, TH, KR, AE} & \cmark & \makecell[l]{Jiripbot, PintSized,\\ Hacktools} & Apr '12 - Jun '15 & V3, V6 & \makecell[l]{S1, S2, S3, S6} \\ \midrule

2015 & -- & IL, PS & -- & DownExecute, Xtreme RAT, Poison Ivy & Jun '14 - Apr '15 & V1 & S1 \\ \midrule

2016 & Lazarus & KR & \xmark & \makecell[l]{OnionDog, Icefog backdoor,\\ USB worms, dropper trojans} & -- & \makecell[l]{V1, V7, V10} & S3, S6 \\ \midrule

2016 & Scarlet Mimic & RU, IN & \xmark & FakeM RAT, MobileOrder, CallMe, SFX archives & -- & \makecell[l]{V1, V7, V6, V3} & \makecell[l]{S1, S7, S10} \\ \midrule

2016 & NetTraveler & \makecell[l]{RU, MN, BY, TR, UA} & \xmark & \makecell[l]{NetTraveler, Saker, DarkStRat, Gh0st RAT, Netbot,\\ PlugX} & -- & \makecell[l]{V1, V6, V7} & \makecell[l]{S1, S2, S7, S10} \\ \midrule

2016 & StrongPity & \makecell[l]{IT, BE, DZ, CI, MA, \\ FR, TN, NL, CA, TR} & -- & StrongPity, trojanized WinRAR, TrueCrypt & Dec '15 - Sep '16 & V3, V11 & S3, S7 \\ \midrule

2017 & Lazarus & PL & \xmark & Shellcode, encrypted EXE files, cambio, perfmon & Oct '16 - Feb '17 & V8 & S4 \\ \midrule

2017 & APT28 & US & \cmark & DealersChoice framework & -- & \makecell[l]{V1, V6, V7} & S1, S2 \\ \midrule

2017 & APT34 & -- & \xmark & POWRUNER, BONDUPDATER & -- & \makecell[l]{V1, V6, V7} & S1 \\ \midrule

2017 & NewsBeef & SA & -- & Shamoon, StoneDrill & Nov '16 - Jan '17 & \makecell[l]{V1, V4, V5} & S2, S3\\ \midrule

2018 & OilRig & -- & \xmark & OopsIE Trojan, QUADAGENT & Jul '18 - Sep '18 & V1, V7 & S1 \\ \midrule

2018 & Goblin Panda & RU & -- & Sisfader RAT & -- & V7 & S1, S2 \\ \midrule

2018 & Patchwork & IN, PK & \xmark & BADNEWS, EPS & Dec '17 - Mar '18 & V6 & S1, S3 \\ \midrule

2018 & Lazarus & -- & \xmark & Fastcash, EternalBlue & -- & V6 & S4 \\ \midrule

2019 & OceanLotus & \makecell[l]{VN, KH, TH, LA, CN} & \xmark & \makecell[l]{OceanLotus RAT, Dll hijack,\\ macOS backdoors, CocCocUpdate dropper} & Apr '12 - Mar '19 & \makecell[l]{V2, V6, V7} & \makecell[l]{S1, S2, S4, S8, S9} \\ \midrule

2019 & APT28 & KZ & \xmark & Zebrocy & -- & V2, V7 & S2 \\ \midrule

2019 & APT39 & US, EG, ES, SA & \xmark & \makecell[l]{SEAWEED, CACHEMONEY, POWBAT, ANTAK,\\ ASPXSpy, Mimikatz, BLUETORCH, REDTRIP, \\PINKTRIP, BLUETRIP} & Nov '14 - Dec '18 & \makecell[l]{V1, V4, V6, V7} & \makecell[l]{S1, S2, S3} \\ \midrule

2019 & Lazarus & RU & \xmark & \makecell[l]{KEYMARBLE, VBScript, CAB} & -- & V7 & S2, S3 \\ \midrule

2020 & AridViper & IL & \xmark & Augury malware addon & Nov '20 - Dec '20 & \makecell[l]{V2, V5, V7} & S1 \\ \midrule

2020 & Transparent Tribe & IN & \xmark & \makecell[l]{Crimson RAT, Oblique RAT, \\AhMyth android RAT} & -- & \makecell[l]{V2, V5, V7} & S1, S7 \\ \midrule

2020 & Darkhotel & CN, JP & \xmark & Ramsay & -- & \makecell[l]{ V6, V7, V11} & S1, S3 \\ \midrule

2020 & Kimsuky & KR & \xmark & \makecell[l]{DLL hijack} & -- & V1, V7 & S1, S7 \\ \midrule

2021 & Earth Wendigo & TW & \xmark & \makecell[l]{JavaScript and WebSocket backdoors,\\ XSS attack scripts, service worker script,\\ WENDIGOE Trojan, Cobalt Strike} & -- & \makecell[l]{V1, V8, V11} & \makecell[l]{S1, S7, S8} \\ \midrule

2021 & NAIKON & -- & -- & RainyDay, Nebulae, Aria-body & Jun '19 - Mar '21 & V6, V8 & S1 \\ \midrule

2021 & GhostEmperor & -- & \cmark & -- & Jul '20 - Jul '21 & V6 & S1, S2 \\ \midrule

2021 & Water Kappa & JP & \xmark & Cinobi, SHELLOAD & -- & V9, V11 & S4, S7 \\ \midrule

2022 & Bitter APT & BD & \cmark & ZxxZ, Almond RAT & Oct '21 - May '22 & V6, V7 & S1 \\ \midrule

2022 & Molerats APT & PS, TR & \xmark & \makecell[l]{Spark, .NET-based backdoor,\\ malicious RAR files} & Jul '21 - Dec '21 & V1, V7 & \makecell[l]{ S1, S4, S9, S10} \\ \midrule

2022 & APT-Q-43 & PK & -- & VajraSpy RAT & Jun '21 - Feb '22 & V5, V9 & S1, S7 \\ \midrule

2022 & Lazarus & KR & \xmark & PowerShell scripts & Mar '22 - Apr '22 & \makecell[l]{V1, V6, V7} & \makecell[l]{S2, S3, S4, S7} \\ \midrule

2023 & APT36 & IN & -- & ElizaRAT, LimePad & -- & V9, V11 & S1, S7 \\ \midrule

2023 & AeroBlade & US & \xmark & MS Office documents & Sep '22 - Jul '23 & V1, V7 & S2 \\ \midrule

2023 & APT28 & \makecell[l]{DE, NL, UK, US, UA} & \xmark & Jaguar Tooth & -- & V6 & S1, S2 \\ \midrule

2023 & APT-C-36 & CO, EC & \xmark & Quasar RAT, Meterpreter & -- & \makecell[l]{V1, V7, V9} & \makecell[l]{S1, S2, S7} \\
\bottomrule

    \end{tabular}
    }
    \label{tab:Appendix_manualAnswers}
\end{table*}

\end{document}